\documentclass[a4paper,11pt]{article}

\usepackage{jheppub} 

\usepackage[T1]{fontenc} 

\usepackage{xcolor}
\usepackage[utf8]{inputenc}
\usepackage{cleveref}
\usepackage{amsmath}
\usepackage{amssymb}
\usepackage{pifont}
\usepackage{graphicx}
\usepackage{multirow}
\usepackage[normalem]{ulem}
\usepackage{comment}
\usepackage{booktabs}
\usepackage{slashed}
\usepackage{makecell}

\usepackage{tikz-feynman}
\tikzfeynmanset{compat=1.1.0}

\newcommand{\braket}[1]{\ensuremath{\left\langle #1 \right\rangle}}

\def\r {\rightarrow}

\newdimen\arrowsize
\newdimen\mylw
\pgfkeys{/my arrows/chemeq/.style={draw,thick,double distance=2pt,onearc-onearc}}
\pgfkeys{/my arrows/size/.code={\pgfsetarrowoptions{onearc}{#1}}}
\def\myalw{.4pt}
\pgfarrowsdeclare{onearc}{onearc}{%
  \mylw=\myalw
  \pgfarrowsleftextend{-\pgfgetarrowoptions{onearc}-.5\mylw}
  \pgfarrowsrightextend{0pt}
}{%
  \pgfsetdash{}{0pt}
  \mylw=\pgflinewidth
  \pgfsetlinewidth{\myalw}
  \advance\arrowsize by.5\pgflinewidth
  \pgfpathmoveto{\pgfpoint{-\pgfgetarrowoptions{onearc}}{-\pgfgetarrowoptions{onearc}-.5\mylw}}%
  \pgfpatharc{180}{90}{\pgfgetarrowoptions{onearc}}
  \pgfusepathqstroke
}

\title{\boldmath  
	 Neutrino masses from new seesaw models:
  Low-scale variants and phenomenological implications}
	 
\author[a,b]{Alessio Giarnetti,}
\author[c,d]{Juan Herrero-García,}
\author[a,b]{Simone Marciano,}
\author[a,b]{Davide Meloni,}
\author[c,d]{and Drona Vatsyayan}

\affiliation[a]{Dipartimento di Matematica e Fisica, Universit\'a di Roma Tre,
Via della Vasca Navale 84, 00146, Roma, Italy}

\affiliation[b]{INFN Sezione di Roma Tre, Via della Vasca Navale 84, 00146, Roma, Italy}

\affiliation[c]{Departament de Física Teòrica, Universitat de València, 46100 Burjassot, Spain}

\affiliation[d]{Instituto de Física Corpuscular (CSIC-Universitat de València),
Parc Científic UV, C/Catedrático José Beltrán, 2, E-46980 Paterna, Spain}

\emailAdd{alessio.giarnetti@uniroma3.it}
\emailAdd{juan.herrero@ific.uv.es}
\emailAdd{simone.marciano@uniroma3.it}
\emailAdd{davide.meloni@uniroma3.it}
\emailAdd{drona.vatsyayan@ific.uv.es}

\abstract{
With just the Standard Model Higgs doublet, there are only three types of seesaw models that generate light Majorana
neutrino masses at tree level after electroweak spontaneous symmetry breaking. However, if there exist additional TeV scalars acquiring vacuum expectation values, coupled with heavier fermionic multiplets, several new seesaw models become possible. These new seesaws are the primary focus of this study and correspond to the tree-level ultraviolet completions of the effective operators studied in a companion publication. We are interested in the genuine cases, in which the standard seesaw contributions are absent. In addition to the tree-level generation of neutrino masses, we also consider the one-loop contributions. Furthermore, we construct low-energy versions that exhibit a very rich phenomenology. Specifically, we scrutinise the generation of dimension-6 operators and explore their implications, including non-unitarity of the leptonic mixing matrix, non-universal $Z-$boson interactions, and lepton flavor violation. Finally, we provide (Generalised) Scotogenic-like variants that incorporate viable dark matter candidates.
}


\preprint{IFIC-23/55}

\makeatletter
\gdef\@fpheader{}
\makeatother

\begin{document} 
\maketitle
\flushbottom

\section{Introduction} \label{sec:intro}

The origin of neutrino masses is one of the open problems of the Standard Model (SM) of particle physics. Arguably, one of the best-motivated ways to generate the smallness of neutrino masses is the seesaws, where neutrinos are Majorana particles, with their mass suppressed by a heavy scale. With just the SM Higgs doublet, only 3 possible mediators exist: a fermion singlet with no hypercharge (Type-I seesaw \cite{Minkowski:1977sc,Yanagida:1980xy,Gell-Mann:1979vob,Mohapatra:1979ia}), a scalar triplet with hypercharge one (Type-II seesaw \cite{Schechter:1980gr,Schechter:1981cv,Lazarides:1980nt,Mohapatra:1980yp}), and a fermion triplet with no hypercharge (Type-III seesaw \cite{Foot:1988aq}). These scenarios, specially Type-I seesaw, are very well-motivated by Grand Unified Theories \cite{Georgi:1974sy,Pati:1974yy,Fritzsch:1974nn} and by the generation of the baryon asymmetry through leptogenesis \cite{Fukugita:1986hr}. However, they are very difficult to test and present a problem of hierarchies \cite{Vissani:1997ys, Casas:2004gh, Herrero-Garcia:2019czj,Arcadi:2022ojj}.

In this work, we study new seesaw models constructed with extra scalar multiplets that take a small induced Vacuum Expectation Value (VEV) and new fermion mediators, which may be either Majorana o vector-like. These models arise as UV completions of new Weinberg-like operators involving scalar multiplets upto the quintuplet representation, studied in a companion publication \cite{Giarnetti:2023dcr}, where the scalar phenomenology is also studied in detail. We focus on those models that are \emph{genuine}, such that the dominant contribution to neutrino masses is proportional at least to the VEV of one new scalar. Otherwise, in the absence of large hierarchies in the Yukawa couplings and/or \emph{ad-hoc} symmetries that we do not want to consider, the contribution to neutrino masses via the SM Higgs VEVs would dominate. Therefore, we do not consider scenarios that involve the usual seesaw contributions.

Their contribution to the Electroweak Precision Tests (EWPTs) makes these scenarios much more testable than the usual high-scale seesaws. Moreover, low-energy variants with the fermions at TeV scale may be constructed. They give rise to a rich phenomenology such as Lepton Flavour Violation (LFV), $Z-$boson mediated Flavour Changing Neutral Currents (FCNC) and universality violation. Related works involving large $SU(2)$ multiplets have been done in the literature, for example, models with scalar quadruplets were studied in Refs.~\cite{Kumericki:2012bh,Picek:2012ei,Babu:2009aq,Bambhaniya:2013yca,Ghosh:2016lnu,Ghosh:2017jbw,Ghosh:2018drw,Dorsner:2019vgf,Dorsner:2021qwg,Picek:2009is,Kumericki:2011hf}, whereas a model with a scalar triplet and a quintuplet was presented in Ref.~\cite{McDonald:2013hsa}. A catalogue of models for neutrino masses involving new scalar multiplets can be found in Refs.~\cite{Bonnet:2009ej,McDonald:2013kca,Wang:2016lve,Anamiati:2018cuq,Law:2013gma,Giarnetti:2023dcr}.

The rest of the paper is structured as follows. In Section~\ref{sec:seesaws} we consider neutrino masses generated by new seesaw models which contain one or two new extra scalar multiplets and either new Majorana or vector-like fermions. In Section~\ref{sec:mnuloop} we study the generation of neutrino masses at one loop. The considered models also generate dimension-$6$ operators at tree level, discussed in Section~\ref{sec:dim6}. In Section~\ref{sec:pheno} we study the phenomenology, including non-unitarity effects and lepton flavour violating processes. In Section~\ref{sec:scot} we outline the possible construction of Scotogenic-like models, with viable dark matter (DM) candidates. Finally, we provide some conclusions in Section~\ref{sec:conc} and also a few appendices with additional material. 

\section{Neutrino masses from new seesaw models} \label{sec:seesaws}

In this section, we want to systematically study the possible ultraviolet (UV) complete models which could provide neutrino masses, involving new scalar multiplets and a heavy mediator. The Effective Field Theory (EFT) approach to this framework has been studied in Ref.~\cite{Giarnetti:2023dcr}. First, we will discuss the three usual seesaws which can be built using the SM Higgs doublet; then, we will explore the possible models in which we add only one or two new scalar multiplets. We restrict to models with at most $SU(2)$ quintuplet representations for the scalars/fermions as problems with unitarity and non-perturbativity arise for larger representations with masses close to the EW scale, due to their large Renormalisation Group Evolution (RGE) running \cite{Hally:2012pu,Earl:2013jsa}.

\subsection{The standard seesaw models}

Let us review the case of just the SM Higgs doublet, $H$. In the SM, up to $SU(2)$ contractions there is a unique dimension-$5$ operator, i.e., the Weinberg operator $LLHH$ \cite{Weinberg:1979sa}. In order to get a $SU(2)$ singlet for the dimension-$5$ Weinberg operator, the possible contractions of the fields are
\begin{equation}
\begin{aligned}
\mathcal{O}^{(0)}_{5,a}=(HL)_\mathbf{1}(HL)_\mathbf{1}\,,\quad &\mathcal{O}^{(0)}_{5,b}=(HL)_\mathbf{3}(HL)_\mathbf{3}\,,\\
\mathcal{O}^{(0)}_{5,c}=(HH)_\mathbf{3}(LL)_\mathbf{3}\,,\quad &\mathcal{O}^{(0)}_{5,d}=(HH)_\mathbf{1}(LL)_\mathbf{1}\,.
\end{aligned}\label{smops}
\end{equation}
Since the singlet $(HH)_\mathbf{1}=0$, the last operator vanishes identically. For the other three cases in Eq.~\eqref{smops}, one can obtain the operator from a UV completion of the SM in which some heavy degrees of freedom are integrated out at tree level. The properties of such mediators are determined by the $SU(2)$ group theory decomposition $\mathbf{2}\otimes \mathbf{2} = \mathbf{1} \oplus \mathbf{3}$, while their nature depends on the particles entering the vertex, which can be of the form fermion-fermion-scalar or scalar-scalar-scalar. As for the hypercharge $Y$, they are fixed as follows. For heavy fermion mediators $\Psi$, the Yukawa interactions of the full theory are of the form
\begin{equation}
y\overline{{\Psi}}\tilde{H} ^\dagger L\,+ {\rm H.c.}\,,
\end{equation}
while for a heavy scalar mediator $S$, the relevant vertex is
\begin{equation}
f\overline{L^c}S L \,+ {\rm H.c.}\,.
\end{equation}
\begin{table}[!htb]
\centering
\begin{tabular}{|l|c|c|c|c|}\hline
& Mediator & $SU(2)$ & $Y$ & Seesaw \\ \hline
$\mathcal{O}^{(0)}_{5,a}$ & fermion $N$& 1 & 0 & Type-I\\
$\mathcal{O}^{(0)}_{5,b}$ & fermion $\Sigma$ & 3 & 0& Type-III\\
$\mathcal{O}^{(0)}_{5,c}$ & scalar $\Delta$& 3 & 1 & Type-II\\ \hline
\end{tabular}
\caption{Summary of relevant properties of the mediators of tree-level UV completions of the SM Weinberg operator (i.e., the seesaws), with their possible $SU(2)$ contractions $\mathcal{O}^{(0)}_{5,a}$, $\mathcal{O}^{(0)}_{5,b}$ and $\mathcal{O}^{(0)}_{5,c}$.}
\label{tab:nsi global}
\end{table}
In Table~\ref{tab:nsi global}, we outline the nature and transformation properties of the mediators involved in the UV completion of the SM Weinberg operator. As it is well known, the UV completions include a hypercharge-less fermion singlet, $N_R$, a $Y=1$ scalar triplet, $\Delta$, and a hypercharge-less fermion triplet, $\Sigma$. These three cases are the well-known Type-I, -II and -III seesaw models, respectively.

\subsection{Ultraviolet completions with new scalar multiplets} \label{sec:extensions}

The possible Weinberg-like operators with up to two new scalars $\Phi_1, \Phi_2$ are: 
\begin{align} 
-\mathcal{L}_5 &=\frac{1}{2}\sum_i {C^{(i)}_5}\,\mathcal{O}^{(i)}_5 + {\rm H.c.}  \nonumber \
\end{align}
where
\begin{equation}\label{eq:newops}
\begin{aligned}
\mathcal{O}^{(0)}_{5}=(LH)_{\mathbf{1}}(LH)_{\mathbf{1}}\,,\qquad &\mathcal{O}^{(1)}_{5}=(LH)_\mathbf{N}(L\Phi_1)_\mathbf{N}\,,\\
\mathcal{O}^{(2)}_{5}=(L\Phi_1)_\mathbf{N} (L\Phi_1)_\mathbf{N}\,,\qquad &\mathcal{O}^{(3)}_{5}=(L\Phi_1)_\mathbf{N} (L\Phi_2)_\mathbf{N}\,,
\end{aligned}
\end{equation}
and $C^{(i)}$ are the dimensionful Wilson coefficients (WCs). Note that $C^{(0)}$ and $C^{(2)}$ are symmetric matrices in flavour space, i.e. $C^{(0)T}=C^{(0)}$ and $C^{(2)T}=C^{(2)}$. The operators in Eq.~(\ref{eq:newops}) may be obtained at low energy by integrating out a heavy mediator at tree level. We have also chosen a \textit{fermion-like} contraction for our basis of operators in Eq.~\eqref{eq:newops}, where $\mathbf{N}$ denotes the highest $SU(2)$ representation of the UV completion. Note that the effective operators in Eq.~(\ref{eq:newops}) have to be understood as classes of operators, encoding all the allowed possible \emph{conjugations} of the scalar fields. Moreover, the other $SU(2)$ contractions of the operators are related to these by Fierz identities. For instance, $\mathcal{O}_{5,b}^{(0)}=-\mathcal{O}_{5,a}^{(0)} = -\frac{1}{2} \mathcal{O}_{5,c}^{(0)}$ (see Eq.~\eqref{smops}).

\subsubsection{One new scalar multiplet} \label{sec:extensions_one}

First, we will study the possibility to include only one new scalar field $\Phi_1$. As already mentioned, we want to consider UV complete {\it genuine} models, namely those in which the heavy mediator is not in common with the three standard seesaws ones.

In Ref.~\cite{Giarnetti:2023dcr}, it has been shown that given the possible contractions of the Weinberg-like operators, in addition to the widely studied 2HDM \cite{Oliver:2001eg,Hernandez-Garcia:2019uof}, the only possible multiplet $\Phi_1$ which could provide neutrino masses is a scalar quadruplet (given our quintuplet cutoff). Morever, regardless of the quantum numbers of the new scalar multiplet, the only scalar mediator allowed by gauge invariance is either a singlet or a triplet. The former does not provide a neutrino mass term, while the latter is identical to the Type-II seesaw one, thus providing a sub-leading contribution. Therefore, the only interesting UV models are those that contain a heavy fermion mediator. 

Considering a Majorana ($Y=0$) mediator $\Sigma$, the Yukawa and fermion mass terms of the Lagrangian can be written as
\begin{equation}
    \mathcal{L}\supset -\overline{L} y_H\widetilde{H} \Sigma-\overline{L}y_1\Phi_1 \Sigma -\dfrac{1}{2}\,\overline{\Sigma^c} M_\Sigma \Sigma + {\rm H.c.} \,,
\end{equation}
where $y_H$ and $y_1$ are the Yukawa matrices of the SM Higgs doublet $H$ and the new scalar field $\Phi_1$, respectively. The first term is invariant only if $\Sigma$ is a triplet or a singlet; the second term is invariant if $\Sigma$ is either a triplet or a quintuplet. If we want to forbid the first term and allow the second one (in order to avoid a dominant contribution from the SM Higgs), the only choice is that $\Sigma$ is a Majorana fermion quintuplet with zero hypercharge, namely $\Sigma=(5,0) \equiv \mathbf{5}_0^F$. 

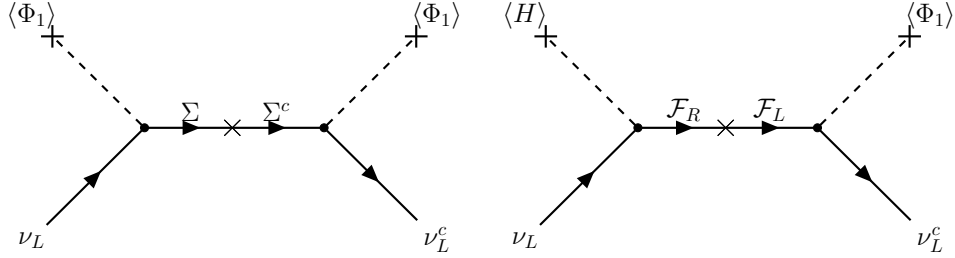
\begin{figure}[!htb]
\centering
\begin{tikzpicture}[scale=0.7, transform shape,every text node part/.style={align=center}]
    \begin{feynman}
        \node [dot] (a);
        \vertex [above left=3cmof a] (b) {\Large{$\braket{\Phi_1}$}};
         \vertex [below left=3cmof a] (c) {\Large{$\nu_L$}};
        \vertex [right=1.65cmof a] (d);
        \node [right=1.65cmof d] [dot] (g);
        \vertex [above right=3cmof g] (e) {\Large{$\braket{\Phi_1}$}};
        \vertex [below right=3cmof g] (f) {\Large{$\nu_L^c$}};
        \diagram* {
            (c) -- [fermion,thick] (a) -- [scalar,thick,insertion=1] (b),
            (a) -- [edge label=\Large{$\Sigma$},insertion=1]  (d),
            (a) -- [fermion,thick]  (d),
            (d) -- [edge label=\Large{$\Sigma^c$},insertion=0] (g),
            (d) -- [fermion,thick]  (g),
            (g) -- [fermion,thick] (f),
            (g) -- [scalar,thick,insertion=1] (e),
        };
    \end{feynman}
\end{tikzpicture}
\hspace{1mm}
\begin{tikzpicture}[scale=0.7, transform shape,every text node part/.style={align=center}]
    \begin{feynman}
        \node [dot] (a);
        \vertex [above left=3cmof a] (b) {\Large{$\braket{H}$}};
         \vertex [below left=3cmof a] (c) {\Large{$\nu_L$}};
        \vertex [right=1.65cmof a] (d);
        \node [right=1.65cmof d] [dot] (g);
        \vertex [above right=3cmof g] (e) {\Large{$\braket{\Phi_1}$}};
        \vertex [below right=3cmof g] (f) {\Large{$\nu_L^c$}};
        \diagram* {
            (c) -- [fermion,thick] (a) -- [scalar,thick,insertion=1] (b),
            (a) -- [edge label=\Large{$\mathcal{F}_R$},insertion=1]  (d),
            (a) -- [fermion,thick]  (d),
            (d) -- [edge label=\Large{$\mathcal{F}_L$},insertion=0] (g),
            (d) -- [fermion,thick]  (g),
            (g) -- [fermion,thick] (f),
            (g) -- [scalar,thick,insertion=1] (e),
        };
    \end{feynman}
\end{tikzpicture}
\caption{Tree-level diagram for neutrino masses with one new scalar multiplet $\Phi_1$ and a Majorana fermion mediator $\Sigma$ (vector-like mediator $\mathcal{F}$) in the left (right) panel.} \label{fig:mnu_1S}
\end{figure}

The hypercharge of $\Phi_1$ must be equal to the lepton doublet in order to allow the $y_1$ coupling. Of course, one needs to ensure that the scalar multiplet contains a neutral component ($Q_i={I_3}_i + Y =0$, where $Q$ is the charge and $I_3$ is the third component of isospin) since they need to take a VEV in order to generate tree-level neutrino masses. Therefore, if the new scalar multiplet transforms under an even (odd) representation of $SU(2)$, it must carry a fractional (integer) hypercharge such that $|Y| \leq ({\bf N}-1)/2$, for a multiplet of dimension ${\bf N}$. Thus, the quantum numbers for $\Phi_1$ are $(4, -1/2)\equiv\mathbf{4}_{-1/2}^S$. We denote this model as $\bf A_1$, which has the same scalar particle content as the scenario $\mathbf{A_I}$ in Ref.~\cite{Giarnetti:2023dcr}.\footnote{Note that a change of sign for the hypercharge corresponds to a redefinition of the field and does not affect our results.} In the case of a Majorana fermion, neutrino masses are generated by the diagram in the left panel of Fig.~\ref{fig:mnu_1S}, which produces the Weinberg-like operator $\mathcal{O}_5^{(2)}$ once the heavy mediator is integrated out.

Let us now consider a vector-like fermion mediator $\mathcal{F}$, with hypercharge $Y_{\mathcal{F}}\not=0$. We can then distinguish the left and right chiral components, so that they can couple differently to the SM Higgs doublet and the new scalar multiplet. The relevant terms of the Lagrangian read
\begin{equation}
    \mathcal{L}\supset  -\overline{L} y_H H \mathcal{F}_R -\overline{L} y_1 \Phi_1\mathcal{F}_L^c -\overline{\mathcal{F}} M_{\mathcal{F}} \mathcal{F}+{\rm H.c.}\,,
\end{equation}
where the vector-like fermion mediator is $\mathcal{F}=\mathcal{F}_R+\mathcal{F}_L$.
In order to provide neutrino masses, the scalar multiplet $\Phi_1$ must be a quadruplet and the fermion mediator $\mathcal{F}$ must be a triplet; the $y_H$ Yukawa term then implies $Y_\mathcal{F} =-1$. The rest of the terms fix the scalar quadruplet to carry a hypercharge $Y_{\Phi_1} =-3/2$. We denote this model containing $\Phi_1=(4,-3/2)\equiv\mathbf{4}_{-3/2}^S$ and $\mathcal{F}=(3,-1)\equiv\mathbf{3}_{-1}^F$ as model $\bf A_2$ which, regarding the scalar sector, corresponds to scenario $\mathbf{A_{II}}$ in Ref.~\cite{Giarnetti:2023dcr}. In this case, neutrino masses are obtained from the diagram in the right panel of Fig.~\ref{fig:mnu_1S}, which generates $\mathcal{O}^{(1)}_5$ when the mediator is integrated out.\footnote{Given the small VEVs of the new scalar multiplets, in the rest of the manuscript we will neglect the mixing between the new particles and the SM charged leptons.}

\subsubsection{Two new scalar multiplets} \label{sec:extensions_two}

Let us now discuss the case in which we add to the SM, two scalar multiplets, $\Phi_1$ and $\Phi_2$. In this case, in addition to $\mathcal{O}^{(1)}_5$ and $\mathcal{O}^{(2)}_5$, another Weinberg-like operator, $\mathcal{O}^{(3)}_5$ (see Eq.~\ref{eq:newops}), may be generated. We will focus on the possible UV completions containing Majorana and vector-like fermions in the 
next two subsections.  

\subsubsection*{Majorana mediators}

The Yukawa and fermion  mass terms of a model with two new scalar multiplets $\Phi_1, \Phi_2$ and a new hypercharge-less Majorana fermion mediator  $\Sigma$ may be written as
\begin{equation}
    \mathcal{L}\supset  -\overline{L} y_1 \Phi_1 \Sigma- \overline{L} y_2 \Phi_2\Sigma   -\dfrac{1}{2}\,\overline \Sigma M_\Sigma \Sigma^c + {\rm H.c.} \,,
\end{equation}
where $y_{1,2}$ are Yukawa couplings.
This Lagrangian structure implies that $Y_{\Phi_1}=Y_{\Phi_2}=-1/2$. As both $\Phi_1$ and $\Phi_2$ have fractional hypercharges, the $SU(2)$ representations of the new multiplets must be even: $\Phi_1=(\mathbf{2X_1}, -1/2)$ and $\Phi_2=(\mathbf{2X_2}, -1/2)$, with $(\mathbf{X_1},\mathbf{X_2})>1$. For the sake of simplicity, we further assume $\mathbf{X_1}\not=\mathbf{X_2}$ to avoid scenarios involving particles with identical quantum numbers, which leads to the same phenomenology as that of a single multiplet discussed above. Let us now come back to the Weinberg-like operator $(L\Phi_1)(L\Phi_2)$ in order to understand the possible singlet contractions. We have:
\begin{equation}
\begin{aligned}
(\Phi_1L)=&\mathbf{2X_1}\otimes\mathbf{2}=\mathbf{2X_1-1}\oplus \mathbf{2X_1+1}\,,\\
(\Phi_2L)=&\mathbf{2X_2}\otimes \mathbf{2}=\mathbf{2X_2-1}\oplus \mathbf{2X_2+1}\,.
\end{aligned}
\end{equation}
Since we need a singlet from the $(L\Phi_1)(L\Phi_2)$  contraction, it follows that $\mathbf{X_2}=\mathbf{X_1+1}$ (taking $\Phi_2$ to be the scalar in the highest representation), and therefore $
{\bf{N_2}}={\bf{N_1}}+2$, where ${\bf{N_i}}=2X_i$ is the (even) dimension of the representation. Thus, in order to have a Majorana mediator, the two representations of the new scalar multiplets must not only be even, but also consecutive. Moreover, to allow both Yukawa interaction terms, the only viable representation for the fermion mediator is $\mathbf{2{N}_1+1}$. The two models with the lowest $SU(2)$ representations in this case are therefore: \emph{i)} a doublet and a quadruplet, and \emph{ii)} a quadruplet and a sextuplet. However, in case \emph{i)} the doublet can be identified with the SM Higgs and we obtain the $\mathbf{A_1}$ model, and regarding \emph{ii)}, we do not study it further as we only consider up to quintuplet representations as discussed above.

\subsubsection*{Vector-like mediators}

We now consider the case in which the new mediator is a vector-like fermion $\mathcal{F}$. In order to avoid the cases already described in the previous sections, we want that the right and left components of the mediator couple to each one of the new scalar multiplets. Thus, the relevant terms in the Lagrangian are
\begin{equation}
\mathcal{L}\supset -\overline{L}y_1\Phi_1 \mathcal{F}_R-\overline{L}y_2\Phi_2 \mathcal{F}_L^c- \overline{\mathcal{F}} M_\mathcal{F} \mathcal{F} +  {\rm H.c.}\,.
\label{eq:lagrVL}
\end{equation}
The group theoretical considerations discussed above hold for the vector-like case as well. However, the key difference is that we do not require the hypercharges of the new scalars or fermions to be equal to $1/2$. Indeed, here we only need that the sum of the hypercharges of the new scalars is $|1|$. In this case, as they are no longer forced to be fractional, $SU(2)$ odd representations are also allowed. Thus, in order to generate $(L\Phi_1)(L\Phi_2)$, we need that the two $SU(2)$ scalar fields are either identical or consecutive even/odd representations. The quantum numbers of the  vector-like fermion mediator are then fixed by the first two terms of Eq.~\eqref{eq:lagrVL}. In particular, when the representations of the two scalar fields are identical, i.e., $\mathbf{N_1}=\mathbf{N_2}=\mathbf{N}$,  the viable representations for the vector-like fermion mediators are $\mathbf{N\pm1}$; on the other hand, if $\mathbf{N_1}<\mathbf{N_2}$, then the mediator must be in the $\mathbf{N_1+1}$ representation. The hypercharge $Y_\mathcal{F}$ is fixed by the relation $Y_\mathcal{F}+Y_{\phi_1}=-1/2$.

\subsection{List of \emph{genuine} models}

Here we list all the possible models with the addition of one or two new scalar multiplets to the SM that generate neutrino masses at tree level via new seesaws. As before, we consider multiplets whose $SU(2)$ representations are at most quintuplets. We are only interested in those \emph{genuine} models that do not generate the Weinberg operator with just SM Higgs doublets, $\mathcal{O}^{(0)}_5$, i.e., that do not generate the usual seesaw contributions. As discussed above, if this was the case, as the SM Higgs VEV is much larger than the new multiplets' VEVs, the usual SM Higgs contribution would be dominant unless strong hierarchies between the involved couplings suppressed it, or some ad-hoc symmetry forbid it completely.\footnote{In Ref.~\cite{Giarnetti:2023dcr} it was shown that there exists two scenarios ($\mathbf{B_V}$ and $\mathbf{B_{VI}}$) containing a triplet scalar with hypercharge -1, from which new Weinberg-like operator might be built. In order to avoid the $\mathcal{O}^{(0)}_5$ Type-II seesaw contribution to be dominant, one needs a strongly suppressed coupling between the lepton doublets and the scalar triplet.}
\begin{table}[!htb]
\centering
\resizebox{\linewidth}{!}{
\begin{tabular}{|c|c|c|c|c|c|c|}
\hline
\textbf{Model}& \textbf{Scalar Multiplets} & \textbf{ Mediators} & {\bf Op.} & {\bf Wilson Coefficients} &\textbf{Refs.} \\ 
\hline\hline
$\bf A_1$&$\Phi_1=\mathbf{4}_{-1/2}^S$& $\Sigma=\mathbf{5}_0^F $&$\mathcal{O}^{(2)}_5$& $C^{(2)}_5=y_1 M_{\Sigma}^{-1} y_1^{T}$& \cite{Kumericki:2012bh,Picek:2012ei}\\ 
\hline
$\bf A_{2}$&$\Phi_1=\mathbf{4}_{-3/2}^S$& $\mathcal{F}=\mathbf{3}_{-1}^F$ &$\mathcal{O}^{(1)}_5$& $C^{(1)}_5=y_1 M_\mathcal{F}^{-1}y_H^{T}+ y_H M_\mathcal{F}^{-1} y_1^T$&\cite{Babu:2009aq,Bambhaniya:2013yca,Ghosh:2016lnu,Ghosh:2017jbw,Ghosh:2018drw,Dorsner:2019vgf,Dorsner:2021qwg}\\ 
\hline
$\bf B_1$&$\Phi_1=\mathbf{4}_{1/2}^S,\;\;\Phi_2=\mathbf{4}_{-3/2}^S$&$\mathcal{F}=\mathbf{5}_{-1}^F$&$\mathcal{O}^{(3)}_5$&$C^{(3)}_5=y_1M_\mathcal{F}^{-1} y_2^{T}+y_2M_\mathcal{F}^{-1} y_1^T$&\cite{Picek:2009is,Kumericki:2011hf}\\ 
\hline
$\bf B_2$&$\Phi_1=\mathbf{3}_0^S,\;\;\Phi_2=\mathbf{5}_{-1}^S$&$\mathcal{F}=\mathbf{4}_{-1/2}^F$&$\mathcal{O}^{(3)}_5$&$C^{(3)}_5=y_1M_\mathcal{F}^{-1} y_2^{T}+y_2 M_\mathcal{F}^{-1} y_1^T$&\cite{Chen:2012vm,McDonald:2013hsa}\\ 
\hline
$\bf B_3$&$\Phi_1=\mathbf{5}_{-2}^S,\;\;\Phi_2=\mathbf{5}_1^S$&$\mathcal{F}=\mathbf{4}_{3/2}^F$&$\mathcal{O}^{(3)}_5$&$C^{(3)}_5=y_1 M_\mathcal{F}^{-1} y_2^{T}+y_2 M_\mathcal{F}^{-1}y_1^T$&$-$\\
\hline
$\bf B_4$&$\Phi_1=\mathbf{5}_{-1}^S,\;\;\Phi_2=\mathbf{5}_0^S$&$\mathcal{F}=\mathbf{4}_{1/2}^F$&$\mathcal{O}^{(3)}_5$&$C^{(3)}_5=y_1 M_\mathcal{F}^{-1} y_2^{T}+y_2 M_\mathcal{F}^{-1} y_1^T$&$-$\\ 
\hline
\end{tabular}
}
\caption{List of \emph{genuine} seesaw models which generate neutrino masses at tree level once the new scalars take VEVs. We show the transformation of the new scalar and fermion particles as ${\bf{N}}^{S,F}_{Y}$, where ${\bf{N}}=2I+1$ is the dimension of the $SU(2)$ representation of weak isospin $I$ and $Y$ is the hypercharge. Models $\bf A_i$ ($\bf B_i$) include one (two) new scalar multiplets. The heavy fermion mediator $\Psi=\Sigma$ ($\mathcal{F}$) is Majorana (vector-like). The fourth column  shows the \emph{Weinberg-like} operator generated, and the fifth one its Wilson Coefficient. References where these models have been studied are listed in the last column. } \label{tab:list} 
\end{table}

In Table~\ref{tab:list} we summarise the \emph{genuine} models, outlining the transformation properties of the new scalar ($S$) and fermion ($F$) mediators as $\Phi={\bf{N}}^{S}_{Y}$ and $\Sigma,\mathcal{F}={\bf{N}}^{F}_{Y}$,  where ${\bf N}=2I+1$ is the dimension of the $SU(2)$ representation of weak isospin $I$ and $Y$ is the hypercharge. 
We also present the Wilson Coefficients associated with the Weinberg-like operators, written in full glory as (with a fermion mediator in the $SU(2)$ representation ${\bf{N}}$) $$ (\overline{L}\Phi_i)_{\mathbf N} ( \Phi_j^T L^c)_{\mathbf N} \; ,$$ where the different contractions are related by Fierz identities, as already mentioned.
The first class of models, denoted by $\bf A_i$, refer to those models in which only one new scalar field is added, while the ${\bf B_i}$ models are those with two new scalars. The fermion mediator $\Psi=\Sigma$ ($\mathcal{F}$) is a Majorana (vector-like) particle. We also list, to the best of our knowledge, references in which similar models have been studied in the literature. It is interesting to note that given the representation cutoff at \textbf{5} for the scalars, the only model in which a Majorana mediator is allowed is the model $\bf A_1$, which only includes one quadruplet. Notice that the models $\mathbf{A_{1,2}}$ and $\mathbf{B_{1-4}}$ correspond in the scalar sector to the scenarios $\mathbf{A_{I,II}}$ and $\mathbf{B_{I-IV}}$ of Ref.~\cite{Giarnetti:2023dcr}, respectively.

After the new scalars take VEVs, neutrino masses will be generated at tree level and they can be written as
\begin{align}
    (m_{\nu})_{\alpha\beta}&=\omega v_1^2 \left(y_1 M_\Sigma^{-1}y_1^T\right)_{\alpha\beta} \qquad &\mathrm{for \, \, \, {\bf A_1}}\,, \label{eq:mnutree1}\\
     (m_{\nu})_{\alpha\beta}&=\omega v_1 v\left(y_H M_{\mathcal{F}}^{-1}y_1^T+y_1 M_{\mathcal{F}}^{-1}y_H^T\right)_{\alpha\beta}\qquad &\mathrm{for \, \, \, {\bf A_2}}\,, \label{eq:mnutree2}\\
     (m_{\nu})_{\alpha\beta}&=\omega v_1 v_2\left(y_1 M_{\mathcal{F}}^{-1}y_2^T+y_2 M_{\mathcal{F}}^{-1}y_1^T\right)_{\alpha\beta} \qquad &\mathrm{for \, \, \, {\bf B_i}}\,, \label{eq:mnutree3}
\end{align}
where $v$, $v_1$ and $v_2$ are the VEVs of the SM Higgs doublet, $\Phi_1$ and $\Phi_2$, respectively, while we denote by
$\omega$ the numerical factor that arises from the contraction of the fields in the three possible dimension-5 operators, $\mathcal{O}^{(1)}_5$, $\mathcal{O}^{(2)}_5$ and $\mathcal{O}^{(3)}_5$, respectively. We summarize the values of these factors in Table~\ref{tab:dim5tab}.

\renewcommand{\arraystretch}{2}
\begin{table}[!htb]
\small
\centering
\begin{tabular}{|c|c|c|c|c|}
\cline{2-5}
 \multicolumn{1}{c |}{} & \textbf{Tree level} & \multicolumn{2}{|c|}{\textbf{Tree level with induced VEVs} }& \textbf{Loop level} \\
 \hline
\cline{1-3} \cline{4-4}
\textbf{Model}& $\omega$& $\xi$&$n$&$\eta$ \\ \hline \hline
$\mathbf{A_1}$                  & $1/2$                          & $1/2\sqrt{3}$ &   $9$  &$-5/6$                 \\ \hline
$\mathbf{A_2}$                  & $-1$                           & $1$           &    $7$  & $2$                    \\ \hline
\multirow{2}{*}{$\mathbf{B_1}$} & \multirow{2}{*}{$-\sqrt{3}/4$} & $1/4$          &  $9$  & \multirow{2}{*}{$5/6$} \\ \cline{3-4}
                                &                                & $-1/12$ $(-1/4)$ & $11$ &                        \\ \hline
$\mathbf{B_2}$                  & $-1/\sqrt{2}$                  & $1/4$           &  $9$ & $5/3$                  \\ \hline
$\mathbf{B_3}$                  & $2$                            & $-1$           &   $7^{\ast}$  & $-5$                   \\ \hline
$\mathbf{B_4}$                  & $-\sqrt{6}$                    & $-3/2$         &   $7^{\ast}$  & $-5$                   \\ \hline
\end{tabular}
\caption{We show the numerical factors $\omega$ ($\xi$) [$\eta$] appearing from the $SU(2)$ contractions in the expression for neutrino masses at tree level (after inducing small VEVs) [at one loop] in the second (third) [fifth] column. We also show in the fourth column the mass dimension $n$ of the generated operator $\mathcal{O}^{(0)}_n=(LH)_{\bf 1} (L H)_{\bf 1} (H^\dagger H)^{\frac{n-5}{2}}$, responsible for the generation of neutrino masses after small VEVs are induced. In model $\mathbf{B}_1$, if neutrino masses come from an $n=11$ operator, the SM Higgs can induce a small VEV for either $\Phi_1$ or $\Phi_2$ (shown in parentheses). The $\ast$ for models $\mathbf{B}_{3,4}$ indicates that the $n=7$ operator cannot be built using only the SM Higgs and it reads  $\mathcal{O}_7=(L\Phi_{i})^2 (H^\dagger H)$. }
\label{tab:dim5tab}
\end{table}
\renewcommand{\arraystretch}{1}

Finally, let us comment on an important distinction between the Majorana and the vector-like type of models. For the case of Majorana  fermions (Model $\bf A_1$), the Yukawa structure  of the mass matrix in Eq.~\eqref{eq:mnutree1} implies that, in  order to have at least two non-zero neutrino masses at tree level, we need at least two copies of the fermion $\Sigma$. On the other hand, for models with vector-like fermions (Models $\bf A_2, B_i$), given the Yukawa structure of the mass matrices in Eqs.~\eqref{eq:mnutree2} and ~\eqref{eq:mnutree3}, having only a single heavy fermion is sufficient to generate two non-zero neutrino masses at tree level. Though in presence of significant loop-corrections to the neutrino mass matrix (see Section~\ref{sec:mnuloop}), a single heavy fermion is enough to generate two non-zero neutrino masses even in the Majorana case \cite{Ren:2011mh}.

\subsection{Neutrino masses and small induced VEVs}

In Section~3 of Ref.~\cite{Giarnetti:2023dcr} the scalar potentials for class-\textbf{A} and \textbf{B} models have been studied in detail and it was shown that the quantum numbers of the multiplets allow writing potential terms which are linear in $\Phi_1$ and/or $\Phi_2$, namely $\lambda_i H^3 \Phi_i,\,  \mu_i H^2 \Phi_i \text{ and }\lambda_{12} H^2 \Phi_1\Phi_2$. The first two terms, for $m_{\Phi_i}\gg v$, can be responsible for a mechanism in which the VEVs of the two new scalar fields can be induced by the SM Higgs VEV and are naturally suppressed with respect to it after the electroweak Spontaneous Symmetry Breaking (SSB), with $v_i\propto \lambda_i v^3/m_{\Phi_i}^2$ or $v_i\propto \mu_i v^2/m_{\Phi_i}^2$. On the other hand, the third {\it mixed} term can suppress one of the two new scalar VEVs with respect to the other: $v_{2(1)}\propto v_{1(2)} v^2/m_{\Phi_{2(1)}}^2$.

This allows to produce neutrino masses from $n>5$ Weinberg-like operators $\mathcal{O}^{(0)}_n=(LH)_{\bf 1} (L H)_{\bf 1} (H^\dagger H)^{\frac{n-5}{2}}$. Let us consider first the $\bf A_i$ models. The effective neutrino mass for model $\bf A_1$ reads in this case
\begin{equation}  \label{eq:A1Higgs}
    (m_{\nu})_{\alpha\beta}=\xi\lambda_{1}^2 \frac{v^6}{4 m_{\Phi_1}^4}\left(y_1M_{\Sigma}^{-1}y_1^T\right)_{\alpha\beta}\,,
\end{equation}
where now $\xi$ is a  numerical factor associated to the contraction of the field in the potential term linear in $\Phi_1$, which is reported in Table~\ref{tab:dim5tab}. The mass dimension of the effective operator in this case is $n=9$. For the $\bf A_2$ model, on the other hand, the neutrino mass matrix reads 
\begin{equation} \label{eq:A2Higgs}
    (m_{\nu})_{\alpha\beta}=\xi\lambda_{1} \frac{v^4}{2m_{\Phi_1}^2} \left(y_H M_{\mathcal{F}}^{-1}y_1^T + y_1 M_{\mathcal{F}}^{-1}y_H^T\right)_{\alpha\beta}\,,
\end{equation}
that corresponds to an $n=7$ operator. 

\begin{figure}[!htb]\centering
    \begin{tikzpicture}[scale=0.8, transform shape,every text node part/.style={align=center}]
        \begin{feynman}
          \node [dot] (a);
            \node [above=1.2cmof a] (bb);
            \node [below left=0.78cmof bb] (name1) {\Large{$\Phi_1$}};
            \node [left=0.3cmof bb] [dot] (b);
            \vertex [right=1.5cmof a] (d);
            \node [right=1.5cmof d] [dot] (g);
            \node [above=1.2cmof g] (ee);
            \node [right=0.3cmof ee] [dot] (e);
            \node [below right=0.78cmof ee] (name2) {\Large{$\Phi_2$}};
            \vertex [below left=2.3cmof a] (c) {\Large{$\nu_L$}};
            \vertex [above=1.5cmof b] (2l) {\Large{$\braket{H}$}};
            \vertex [left=1.1cmof 2l] (1l) {\Large{$\braket{H}$}};
            \vertex [right=1.1cmof 2l] (3l) {\Large{$\braket{H}$}};
            
            \vertex [above=1.5cmof e] (2r) {\Large{$\braket{H}$}};
            \vertex [left=1.1cmof 2r] (1r) {\Large{$\braket{H}$}};
            \vertex [right=1.1cmof 2r] (3r) {\Large{$\braket{H}$}};
            \vertex [below right=2.3cmof g] (f) {\Large{$\nu_L^c$}};
            \diagram* {
                (c) -- [fermion,thick] (a);
                (a) -- [scalar,thick] (b),
                (b) -- [scalar,thick, insertion=1] (1l),
                (b) -- [scalar,thick, insertion=1] (2l),
                (b) -- [scalar,thick, insertion=1] (3l),
                (a) -- [edge label'=\Large{$\mathcal{F}_R$},insertion=1]  (d),
                (a) -- [fermion,thick]  (d),
                (d) -- [edge label'=\Large{$\mathcal{F}_L$},insertion=0] (g),
                (d) -- [fermion,thick]  (g),
                (g) -- [fermion,thick] (f),
                (g) -- [scalar,thick] (e),
                (e) -- [scalar,thick, insertion=1] (1r),
                (e) -- [scalar,thick, insertion=1] (2r),
                (e) -- [scalar,thick, insertion=1] (3r),
            };
            \end{feynman}
    \end{tikzpicture}
    \hspace{3em}
    \begin{tikzpicture}[scale=0.8, transform shape,every text node part/.style={align=center}]
        \begin{feynman}
          \node [dot] (a);
            \node [above=1.2cmof a] (bb);
            \node [below left=0.78cmof bb] (name1) {\Large{$\Phi_1$}};
            \node [left=0.3cmof bb] [dot] (b);
            \vertex [right=1.5cmof a] (d);
            \node [right=1.5cmof d] [dot] (g);
            \node [above=1.2cmof g] (ee);
            \node [below right=0.78cmof ee] (name2) {\Large{$\Phi_2$}};
            \node [right=0.3cmof ee] [dot] (e);
            \node [above left=1.6cmof e] (upright11);
            \node [below=0.6cmof upright11] [dot] (upright1);
            \node [right=1.6cmof e] (right2);
            \vertex [below left=2.3cmof a] (c) {\Large{$\nu_L$}};
            \vertex [above=1.7cmof b] (2l) {\Large{$\braket{H}$}};
            \vertex [left=1.2cmof 2l] (1l) {\Large{$\braket{H}$}};
            \vertex [right=1.2cmof 2l] (3l) {\Large{$\braket{H}$}};
            \vertex [above=1.7cmof upright1] (1r) {\Large{$\braket{H}$}};
            \vertex [right=1cmof 1r] (2r) {\Large{$\braket{H}$}};
            \vertex [below right=0.8cmof 2r] (3r) {\Large{$\braket{H}$}};
            \vertex [above=0.42cmof right2] (right1) {\Large{$\braket{H}$}};
            \vertex [below=0.32cmof right2] (right2) {\Large{$\braket{H}$}};
            \vertex [below right=2.3cmof g] (f) {\Large{$\nu_L^c$}};
            \diagram* {
                (c) -- [fermion,thick] (a);
                (a) -- [scalar,thick] (b),
                (b) -- [scalar,thick, insertion=1] (1l),
                (b) -- [scalar,thick, insertion=1] (2l),
                (b) -- [scalar,thick, insertion=1] (3l),
                (a) -- [edge label'=\Large{$\mathcal{F}_R$},insertion=1]  (d),
                (a) -- [fermion,thick]  (d),
                (d) -- [edge label'=\Large{$\mathcal{F}_L$},insertion=0] (g),
                (d) -- [fermion,thick]  (g),
                (g) -- [fermion,thick] (f),
                (g) -- [scalar,thick] (e),
                (e) -- [scalar,thick, edge label'=\Large{$\Phi_1$}] (upright1),
                (e) -- [scalar,thick,insertion=1] (right1),
                (e) -- [scalar,thick,insertion=1] (right2),
                (upright1) -- [scalar,thick, insertion=1] (1r),
                (upright1) -- [scalar,thick, insertion=1] (2r),
                (upright1) -- [scalar,thick, insertion=1] (3r),
                
            };
            \end{feynman}
    \end{tikzpicture}
    \caption{Example of neutrino mass generation due to induced VEVs. We show the case of Model ${\bf B_1}$, where neutrino masses are generated at tree level via dimension $n=9$ ($n=11$) operators in the left (right) figures.  In addition to the usual Higgs doublet $H$, the model includes the vector-like fermion $\mathbf{5}_{-1}^F$ and the scalar quadruplets: $\mathbf{4}_{1/2}^S$, $\mathbf{4}_{-3/2}^S$.}
    \label{fig:dim>5Diagrams}
    \end{figure}
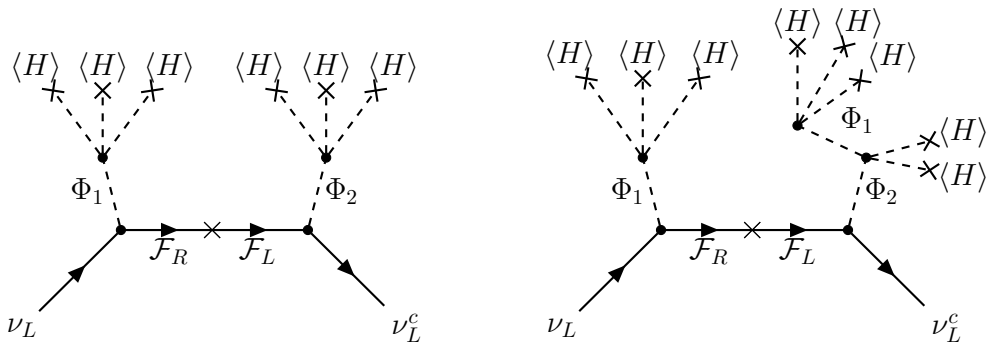

Considering now the more complicated $\bf B_i$ models, for the $\bf B_1$ model, neutrino masses could arise from an $n=9$ or an $n=11$ operator, depending on how the new multiplets' VEVs are induced. If both VEVs are induced directly from the SM Higgs one, neutrino masses arise from the left diagram of Fig.~\ref{fig:dim>5Diagrams}, i.e., a dimension-$9$ operator. They read
\begin{equation} \label{eq:B1Higgs}
    (m_{\nu})_{\alpha\beta}=\xi \lambda_{1}\lambda_{2} \frac{v^6}{4m_{\Phi_1}^2 m_{\Phi_2}^2} \left(y_1M_{\mathcal{F}}^{-1}y_2^T+y_2M_{\mathcal{F}}^{-1}y_1^T\right)_{\alpha\beta}\,.
\end{equation}
On the other hand, if one of the new scalar fields' VEV, say $v_2$, is induced through the mixed potential term $\lambda_{12}H^2\Phi_1\Phi_2$, so that $v_1\gg v_2$, we obtain an $n=11$ operator, see right diagram of Fig.~\ref{fig:dim>5Diagrams}, which leads to the following neutrino masses,
\begin{equation} \label{eq:B1Higgsb}
    (m_{\nu})_{\alpha\beta}=\xi \lambda_{1}^2 \lambda_{12} \frac{v^8}{8m_{\Phi_1}^4 m_{\Phi_2}^2} \left(y_1M_{\mathcal{F}}^{-1}y_2^T+y_2M_{\mathcal{F}}^{-1}y_1^T\right)_{\alpha\beta}\,,
\end{equation}
In the case in which $v_1\ll v_2$, the neutrino mass matrix can be obtained from Eq.~\eqref{eq:B1Higgsb} with the following substitutions: $\lambda_{1}\to\lambda_{2}$ and $m_{\Phi_1}\leftrightarrow\, m_{\Phi_2}$. 

In the $\bf B_2$ model, if we want to suppress both new VEVs, we need to induce the triplet VEV $v_1$ directly from the Higgs VEV $v$, and the quintuplet VEV $v_2$ from $v_1$ through the mixed potential term. Thus, the neutrino mass matrix arises from an $n=9$ operator and reads
\begin{equation} \label{eq:B2Higgs}
    (m_{\nu})_{\alpha\beta}=\xi\mu_{1}^2 \lambda_{12} \frac{v^6}{8m_{\Phi_1}^4 m_{\Phi_2}^2} \left(y_1M_{\mathcal{F}}^{-1}y_2^T+y_2M_{\mathcal{F}}^{-1}y_1^T\right)_{\alpha\beta}\,.
\end{equation}
In the last two models $\bf B_3$ and $\bf B_4$, we do not have a potential term linear in only $\Phi_1$ or $\Phi_2$. Thus, there does not exist a mechanism which ensures that both VEVs are naturally small. However, neutrino masses may still arise from an $n=7$ operator, with say $v_2$ induced by $v_1$ (i.e., $v_2\ll v_1$),
\begin{equation} \label{eq:B3Higgs}
    (m_{\nu})_{\alpha\beta}=\xi \lambda_{12} \frac{v^2}{ 2m_{\Phi_2}^2}v_1^2 \left(y_1M_{\mathcal{F}}^{-1}y_2^T+y_2M_{\mathcal{F}}^{-1}y_1^T\right)_{\alpha\beta}\,.
\end{equation}
In the opposite case, the neutrino mass matrix can be obtained with the substitutions $v_1\to v_2$ and $m_{\Phi_2}\to m_{\Phi_1}$. Notice that, in this case, the $n=7$ Weinberg-like operator is of the form $\mathcal{O}_7=(L\phi_{i})^2 (H^\dagger H)$. 

\subsection{Low-scale \emph{genuine} seesaw models} \label{sec:low}
Although the UV scale of the fermion mediators in the seesaws involving new scalar multiplets is generically lower than the standard seesaw one, due to the suppression of the new VEVs, it is still too large to give rise to low-energy phenomenology involving the fermions in the direct seesaw case. Therefore, in this section we construct low-scale (i.e., TeV) variants of the \emph{genuine} models discussed so far (see Ref.~\cite{Boucenna:2014zba} for a review of low-scale versions of the standard seesaw). Contrary to the high-scale versions, these generate interesting phenomenology: FCNCs, non-unitarity of the PMNS neutrino mixing matrix, and LFV, which will be studied in Section~\ref{sec:pheno}.

\subsubsection{An inverse seesaw version of Model $\bf A_1$} \label{sec:ISS_A1}

Here we construct an inverse seesaw version of the model  $\bf A_1$, which will generate at low energy dimension 5 and 6 operators whose WCs are not strictly related to each other. To this aim, we introduce a quasi-Dirac mediator $\Sigma = \mathbf{5}_{0}^F$, with small Majorana mass terms. The relevant terms in the Lagrangian are as follows:
\begin{eqnarray} \label{eq:ISS_Maj}
\mathcal{L} \supset - y_1 \bar L \Phi_1 \Sigma_R - M  \bar{\Sigma}  \Sigma
- \frac{\mu_R}{2} \bar{\Sigma_R^c}
\Sigma_R  - \frac{\mu_L}{2} \bar{\Sigma_L^c}
\Sigma_L +{\rm H.c.}\,.
\end{eqnarray}
Notice that gauge invariance would also permit a term of the form $\bar L \Phi_1 \Sigma_L^c$, 
which, however, can be rotated away by a redefinition of the field $\Sigma$. After integrating out the $\Sigma$ field,  one obtains dimension 5 and 6 operators with the following WCs, respectively:
\begin{eqnarray}
C^{(2)}_5 &=& -\frac{y_1\,\mu_L\,y_1^T}{2 M^2}\,,\qquad
C^{(1)}_6 = \frac{y_1\,y_1^\dagger}{M^2}\,, 
\end{eqnarray}
from which we clearly see that they are independent of each other, as anticipated.

\subsubsection{Low-scale vector-like models $\bf A_2$, $\bf B_i$} \label{sec:VL_low}

In models with vector-like mediators, lepton number is violated by the product of two different Yukawas and two different VEVs. Therefore, neutrino masses can be reproduced at low scales, say with $M_\mathcal{F} \simeq \mathcal{O}(1)$ TeV, by taking a small enough product of the Yukawas and VEVs. Moreover, if there is a large hierarchy among the Yukawas and/or the VEVs,  some of the Yukawas may be of significant size, generating large contributions to dimension-$6$ operators, see Section~\ref{sec:dim6}. For instance, for Model $\bf A_2$, one may take 
\begin{equation}
    y_1 v_1 \ll y_H v_H \,.
\end{equation}
The other hierarchy is also possible; it requires $y_H \ll 0.01 y_1$. For $y_H \simeq 1$, reproducing neutrino masses with TeV-scale fermion mediators demands
\begin{equation}
    \left(\frac{y_1}{10^{-10}} \right)\left(\frac{v_1}{\text{GeV}} \right) \simeq 1\,.
\end{equation}
If this constraint is realised, processes mediated by $y_H$, like radiative LFV decays, Higgs LFV or $Z$-mediated processes will be significant, see Section~\ref{sec:pheno}.

Regarding the $\bf B_i$ models, neutrino masses can be similarly reproduced with a large hierarchy between the different Yukawas and VEVs, so that there are phenomenological signals, i.e., either
\begin{equation}
  y_1 v_1\ll y_2 v_2\,,
\end{equation}
or equivalently the opposite one. In these models, reproducing neutrino masses with TeV-scale fermion mediators requires
\begin{equation} \label{eq:low_vec}
    \left(\frac{y_1 v_1}{10^{-8}\,\text{GeV}} \right) \left(\frac{y_2 v_2}{\text{GeV}} \right) \simeq 1\,.
\end{equation}
These low-scale versions of the models with vector-like fermionic mediators are the equivalent of inverse seesaw for hypercharge-less fermions. The largest rates for processes generated by dimension-$6$ operators, see Section~\ref{sec:dim6}, are for $y_2 \simeq 1$ and $v_2 \simeq 1$ GeV (the largest it can be from EWPTs). In this case $y_1 v_1  \simeq 10^{-8}\,\text{GeV}$, and the phenomenological signals are due to the second multiplet, $\Phi_2$.

\section{One-loop contributions to neutrino masses} \label{sec:mnuloop}

In the considered models, there may also be a contribution to neutrino masses at the one-loop level, which, as we will see in this section, is not granted to be always subdominant with respect to the tree level contribution. For instance, if the new scalars do not take a VEV, neutrino masses would be generated at loop level only.

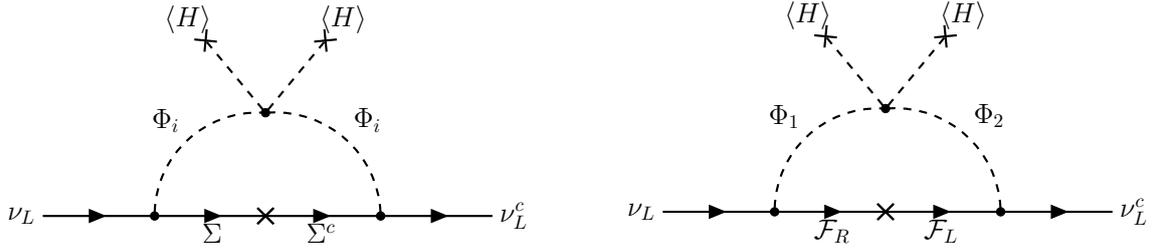
\begin{figure}[!t]
\begin{tikzpicture}[scale=0.73, transform shape,every text node part/.style={align=center}]
    \begin{feynman}
        \node [dot] (b);
            \vertex[left=2.4cmof b] (a) {\Large{$\nu_L$}};
            \vertex [right=2cmof b] (c);
            \node [right=2cmof c] [dot] (d);
            \node [above=1.8cmof c] [dot] (e);
            \node [above=1.7cmof e] (up);
            \vertex [left=1.4cmof up] (upL) {\Large{$\braket{H}$}};
            \vertex [right=1.4cmof up] (upR) {\Large{$\braket{H}$}};
            \vertex [right=2.4cmof d] (f) {\Large{$\nu_L^c$}};
            \diagram* {
                (a) -- [fermion,thick] (b),
                (b) -- [edge label'=\Large{$\Sigma$},insertion=1] (c),
                (b) -- [fermion,thick] (c), 
                (c) -- [fermion,thick,edge label'=\Large{$\Sigma^c$},insertion=0] (d),
                (d) -- [fermion,thick] (f),
                (b) -- [scalar, thick,quarter left,edge label=\Large{$\Phi_i$}] (e),
                (d) -- [scalar,thick,quarter right,edge label'=\Large{$\Phi_i$}] (e),
                (e) -- [scalar,thick,insertion=1] (upL),
                (e) -- [scalar,thick,insertion=1] (upR),
                };
    \end{feynman}
\end{tikzpicture}
\hspace{2.5em}
\begin{tikzpicture}[scale=0.73, transform shape,every text node part/.style={align=center}]
    \begin{feynman}
        \node [dot] (b);
            \vertex[left=2.4cmof b] (a) {\Large{$\nu_L$}};
            \vertex [right=2cmof b] (c);
            \node [right=2cmof c] [dot] (d);
            \node [above=1.8cmof c] [dot] (e);
            \node [above=1.7cmof e] (up);
            \vertex [left=1.4cmof up] (upL) {\Large{$\braket{H}$}};
            \vertex [right=1.4cmof up] (upR) {\Large{$\braket{H}$}};
            \vertex [right=2.4cmof d] (f) {\Large{$\nu_L^c$}};
            \diagram* {
                (a) -- [fermion,thick] (b),
                (b) -- [edge label'=\Large{$\mathcal{F}_R$},insertion=1] (c),
                (b) -- [fermion,thick] (c), 
                (c) -- [fermion,thick,edge label'=\Large{$\mathcal{F}_L$},insertion=0] (d),
                (d) -- [fermion,thick] (f),
                (b) -- [scalar, thick,quarter left,edge label=\Large{$\Phi_1$}] (e),
                (d) -- [scalar,thick,quarter right,edge label'=\Large{$\Phi_2$}] (e),
                (e) -- [scalar,thick,insertion=1] (upL),
                (e) -- [scalar,thick,insertion=1] (upR),
                };
    \end{feynman}
\end{tikzpicture}
\caption{One-loop diagram contributing to Majorana neutrino masses in Model $\bf A_1$ with heavy Majorana mediator (\textit{left}) and models $\bf A_2, B_i$ with heavy vector-like fermions (\textit{right}).} \label{fig:loop}
\end{figure}
Given the transformation properties of the new scalar multiplets, our models always have potential terms which lead to a one-loop diagram proportional to $v^2$. These are $\bar \lambda\Phi_1^2H^2$ for model $\bf A_1$, $\lambda_{1}\Phi_1H^3$ for model $\bf A_2$ and $\lambda_{12}\Phi_1\Phi_2H^2$ for the $\bf B_i$ models. An example of one-loop diagram for $\bf A_1$ (left) and $\bf A_2, B_i$ models (right) is depicted in Fig.~\ref{fig:loop}. This diagram is not the only one contributing to the one-loop neutrino mass correction for our models; however, all the other diagrams give subleading contributions being proportional to at least one of the small new VEVs. 

We want now to evaluate the one-loop contribution to neutrino masses in order to compare them to the tree-level one. We will assume no mixing with SM leptons (therefore no contributions of $H,Z,W$, only heavy particles in the loop, see also Ref.~\cite{Pilaftsis:1991ug}) and degenerate components within the scalars. These one-loop corrections read
\begin{align}
(m_{\nu})^{\rm loop}_{\alpha\beta}&=\eta\,\bar\lambda\,\frac{v^2}{8\pi^2}  \sum_{k}\, y_{1,\alpha k}\, y_{1,\beta k}\, \frac{{M}_\Sigma}{\left(M^2_{(\Phi_1)_0^R}-M^2_{(\Phi_1)_0^I}\right)} \, \, F_2(M_{(\Phi_1)_0^R},M_{(\Phi_1)_0^I},{M}_\Sigma)\quad \mathrm{for \, \, \, {\bf A_1}}\,, \nonumber\\
(m_{\nu})^{\rm loop}_{\alpha\beta}&=\eta \,\lambda_{1}\,\frac{v^2}{8\pi^2}  \left(y_Hy_1^T + y_1y_H^T\right)_{\alpha\beta}\,  \frac{M_\mathcal{F}}{\left(M^2_{\Phi_1}-M^2_{H}\right)}\,F_2(M_{\Phi_1},M_H,M_\mathcal{F})\quad \mathrm{for \, \, \, {\bf A_2}}\,, \nonumber\\
(m_{\nu})^{\rm loop}_{\alpha\beta}&=\eta \,\lambda_{12}\,\frac{v^2}{8\pi^2}  \left(y_1y_2^T+y_2y_1^T\right)_{\alpha\beta}\,  \frac{M_\mathcal{F}}{\left(M^2_{\Phi_1}-M^2_{\Phi_2}\right)} \,F_2(M_{\Phi_1},M_{\Phi_2},M_\mathcal{F}) \quad \mathrm{for \, \, \, {\bf B_i}}\,, \label{eq:mnu1loop}
\end{align}
where $\eta$ are numerical factors (summarised in Table~\ref{tab:dim5tab}) coming from the field contractions, ${M}_\Sigma~(M_\mathcal{F})$ is the mass of the heavy Majorana (vector-like) fermions, $M_{(\Phi_1)_0^R}$ and $M_{(\Phi_1)_0^I}$ are the masses of the neutral CP-even and CP-odd components of $\Phi_1$, respectively, in the Model $\bf A_1$, and $F_2$ is the loop function
\begin{equation}\label{eq:loopmnu}
     F_2(x,y,z)= \frac{x^2}{x^2-z^2}\ln{\frac{x^2}{z^2}}-\frac{y^2}{y^2-z^2}\ln{\frac{y^2}{z^2}} \nonumber \,.
\end{equation}
It is important to mention that both the neutral and the singly-charged scalars can run in the loop.

Let us now compare the contributions to neutrino masses at tree level after the small VEVs induction, Eqs.~\eqref{eq:A1Higgs}-\eqref{eq:B3Higgs}, and at one loop, Eq.~\eqref{eq:mnu1loop}. We will consider the case where the fermions are much heavier than the rest of the scalars, i.e., $M_\Sigma, M_{\mathcal{F}} \gg m_{\Phi_i}$. Neglecting logarithms in Eq.~\eqref{eq:mnu1loop}, and assuming just one copy of the heavy fermion, we obtain
\begin{align}  
m_{\nu}^{\rm loop} &\simeq \,\frac{1}{8\pi^2}\, \left(\frac{\eta\,\bar\lambda}{\xi  \lambda_{1}^2 }\right)\,\left(\frac{4M_{\Phi_1}^4}{v^4}\right)\,m^{\rm tree}_{\nu}\qquad &\mathrm{for \, \, \, {\bf A_1}}\,, \label{eq:compA1} \\ 
m_{\nu}^{\rm loop} &\simeq   \,\frac{1}{8\pi^2}  \left(\frac{ \eta}{\xi}\right)\,\left(\frac{2M_{\Phi_1}^2}{v^2}\right)\,m^{\rm tree}_{\nu}  \qquad &\mathrm{for \, \, \, {\bf A_2}}\,, \label{eq:compA2} \\ 
m_{\nu}^{\rm loop} &\simeq \frac{1}{8\pi^2} \left(
\frac{\eta\,\lambda_{12}}{\xi \lambda_{1}\lambda_{2} } \right)\,\left(\frac{4M_{\Phi_1}^2 M_{\Phi_2}^2}{v^4}  \right)\,m^{\rm tree}_{\nu} 
\qquad &\mathrm{for \, \, \, {\bf B_1} \, \, \mathrm{(n=9)}}\,, \label{eq:compB1}  \\ 
m_{\nu}^{\rm loop} &\simeq \frac{1}{8\pi^2} \left(
\frac{\eta }{\xi \lambda_{1(2)}^2 } \right)\,\left(\frac{8M_{\Phi_{1(2)}}^4 M_{\Phi_{2(1)}}^2}{v^6}  \right)\,m^{\rm tree}_{\nu} 
\qquad &\mathrm{for \, \, \, {\bf B_1} \, \, \mathrm{(n=11)}}\,, \label{eq:compB1bis} \\ 
m_{\nu}^{\rm loop} &\simeq  \frac{1}{8\pi^2}
\left(\frac{\eta}{\xi}\right) \,\left(\frac{8 M_{\Phi_1}^4 M_{\Phi_2}^2} { v^4\mu_{1}^2}\right) \,m^{\rm tree}_{\nu} 
\qquad &\mathrm{for \, \, \, {\bf B_2}}\,, \label{eq:compB2} \\ 
m_{\nu}^{\rm loop}&\simeq \frac{1}{8\pi^2} \left(
\frac{\eta}{\xi}\right)\,\left(\frac{2M_{\Phi_{2{(1)}}}^2}{v_{1(2)}^2} \right) \,m^{\rm tree}_{\nu} 
\qquad &\mathrm{for \, \, \, {\bf B_{3\, (4)}}}\,.  \label{eq:compB34}
\end{align}
Some observations are in order. The mass of the new scalar must be larger than the SM Higgs VEV in order to induce the small VEVs for the new multiplets; this always enhances the loop contribution with respect to the tree level one. 
Assuming in Eq.~(\ref{eq:compB2}) $\mu_1\simeq M_{\Phi_{1}}$, which is the upper limit for the dimensionful parameter $\mu_1$,\footnote{Notice that naturality and charge-breaking constraints imply that $\mu_{1} \lesssim M_{\Phi_1}$, so none of the trilinear couplings can be made arbitrarily large. See for instance, Refs~\cite{Herrero-Garcia:2014hfa,Herrero-Garcia:2017xdu,Casas:1996de}.} there are no couplings in the potential of the models $\bf A_2$, $\bf B_2$, $\bf B_3$ and $\bf B_4$, able to suppress the loop contribution.

In the remaining models, there are some couplings in the potential that control the contributions to neutrino masses. In particular, for Models $\bf A_1$ and $\bf B_1$ ($n=9$), $\bar{\lambda}$ and $\lambda_{12}$ contribute at the loop level, and even if they vanish, the scalars still take small induced VEVs. In the remaining case, namely in $\bf B_1$ ($n=11$), the coupling responsible for the loop level diagram $\lambda_{12}
$ is also involved in the generation of the induced VEVs and gets cancelled in the $m_{\nu}^{\rm loop}/m_{\nu}^{\rm tree}$ ratio. However, in this case, the couplings $\lambda_{1(2)}$ (depending on which of the two VEVs is directly induced by the Higgs VEV) survives in the denominator and can lead to the enhancement of the loop mass with respect to the tree level mass. 
\begin{figure}[!tb]
    \centering
    \includegraphics[width=0.49\linewidth]{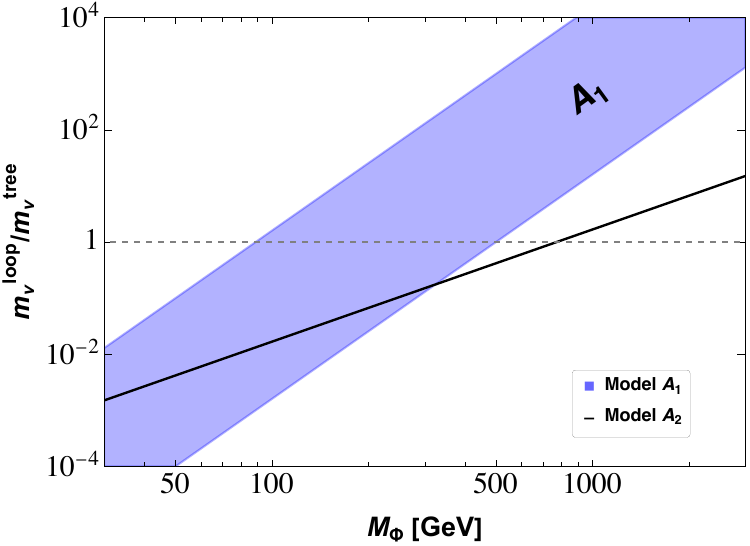}
    \includegraphics[width=0.49\linewidth]{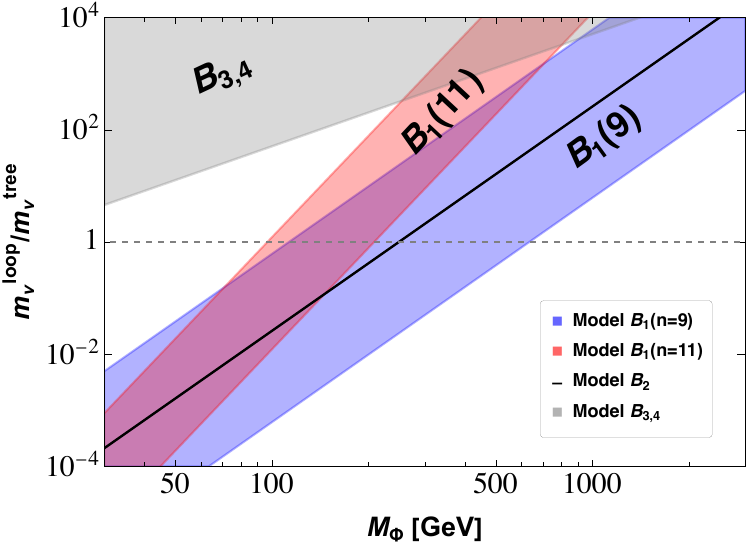}
    \caption{Ratio of the contribution to neutrino masses at one loop and at tree level versus the scalars mass in the limit $M_{\Psi}\gg M_{\Phi}$ for class-\textbf{A} (\textit{left}) and class-\textbf{B} (\textit{right}) models. The colored bands are obtained taking the couplings $\lambda_i\in[0.1;1]$. Notice that the dependence on $\lambda_i$ drops in models $\bf A_1$ and $\bf B_2$. For $\bf B_3$ and $\bf B_4$, the behaviour is very similar, therefore we report only $\bf B_3$. See main text for further details. }
    \label{fig:mtree/mloop}
\end{figure}

In Fig.~\ref{fig:mtree/mloop}, we show how the ratio $m_{\nu}^{\rm loop}/m_{\nu}^{\rm tree}$ depends on the scalar masses, taking all the quartic couplings in the range [0.1, 1], considering Eqs.~\eqref{eq:compA1}-\eqref{eq:compB34}. It is interesting to notice that for the class-$\bf A$ models, the tree-level contribution may dominate considering the most favourable choice of the couplings if $M_{\Phi}<780$ GeV for $\bf A_2$ and $M_{\Phi}<500$ GeV  for $\bf A_1$. In the class-$\bf B$ models, the scalar masses for which the tree level contribution can overcome the loop-level one are smaller in all the cases except for $\bf B_1$ ($n=9$), where we obtain $M_{\Phi}<640$ GeV. In particular, we have $M_{\Phi}<210\,,250$ GeV for $\bf B_1$ ($n=11$) and $\bf B_2$, respectively. For Models $\bf B_{3,4}$, the tree level contribution is always sub-leading because the scalar potential does not allow inducing small VEVs for $\Phi_1$ and $\Phi_2$ and the tree-level contribution is always proportional to either $v_1$ or $v_2$.

In the following, we analyse how the quartic couplings modify the ratio $m_\nu^\text{loop}/m_\nu^{\text{tree}}$ in the $\bf{A_1}$ and $\bf{B_{1}}$ models. Note that the couplings associated with lepton number violation (LNV) generate also a mass for pseudo-Nambu-Goldstone bosons, which should be larger than $m_J^{\rm min}=45$ GeV from $Z-$boson decays \cite{CMS:2022ett}. In Model $\bf{A_1}$, there exists a single massive pseudo-Nambu-Goldstone boson, the pseudo-Majoron $J$; expanding its mass to first order in $\mathcal{O}(v_{1}/v)$, we obtain the lower limit
\begin{equation}
    |\bar \lambda|\geq \dfrac{(m_J^\mathrm{\rm min})^2\,v_{1}}{3\,\sqrt{3}\,v^3}\;\,.
\end{equation}
A small vev $v_1$ is induced after EW SSB, $v_1=\lambda_1 v^3/(2\,M_\Phi^2)$, so that the limit given by the Majoron mass implies 
\begin{equation}
    \frac{\overline{\lambda}}{\lambda_1}\geq \frac{3\,\sqrt{3}(m_J^{\text{min}})^2}{(2\,M_\Phi^2)}\;.
    \label{majoronLimit}
\end{equation}
In Model $\bf{B_1}$, on the other hand, the presence of an explicitly broken additional $U(1)$ symmetry in the scalar potential implies that two massive pseudo-Nambu-Goldstone bosons ($J$ and $\Omega$) appear in the spectrum. Therefore, we obtain the following lower limits: \footnote{Notice that the coupling $\lambda_{12}$ may give subleading contributions to pseudo-Nambu-Goldstone bosons masses proportional to $v^2$, as reported in Ref.~\cite{Giarnetti:2023dcr}. Therefore, $\lambda_{1}$ and $\lambda_2$ bounds could be relaxed if $\lambda_{12}/\lambda_{1,2}\sim\mathcal{O}(v/v_\Phi)$.}
\begin{equation}
    |\lambda_1|\geq \dfrac{2\,\sqrt{3}\,(m_J^\mathrm{min})^2\,v_{1}}{v^3}\;\,,
\end{equation}
\begin{equation}
    |\lambda_2|\geq \dfrac{2\,(m_\Omega^\mathrm{min})\,v_{2}}{v^3}\;\,.
\end{equation}
which are always satisfied for $M_\Phi\gtrsim m_{J,\Omega}^\text{min}$.

\begin{figure}[!htb]
    \centering
    \includegraphics[width=0.5\linewidth]{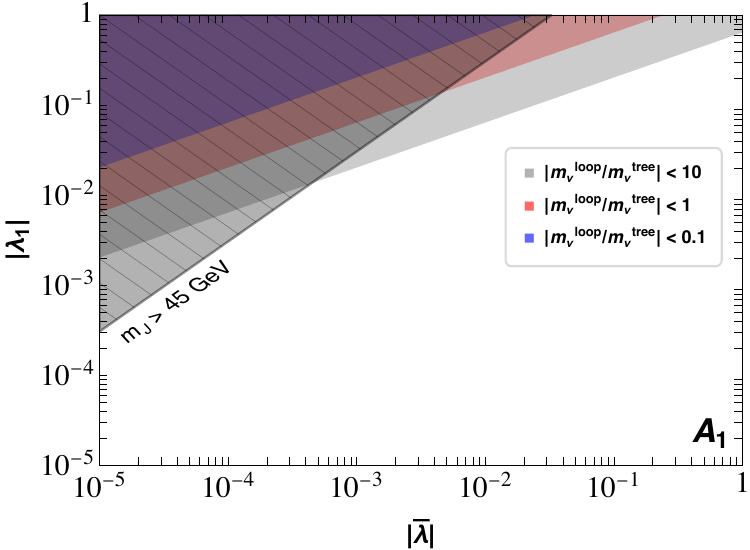}
    \caption{Contours of fixed $m_\nu^\text{loop}/m_\nu^{\text{tree}}$ in the plane of the LNV quartic couplings for Model ${\bf A_1}$. We also show the region excluded by the pseudo-Nambu-Goldstone mass. We take $M_\Phi=400$ GeV. The region with oblique gray lines is excluded by the limit on the Majoron mass given in Eq.~\eqref{majoronLimit}.}
    \label{fig:A1}
\end{figure}

\begin{figure}[!htb]
    \centering
    \includegraphics[width=0.495\linewidth]{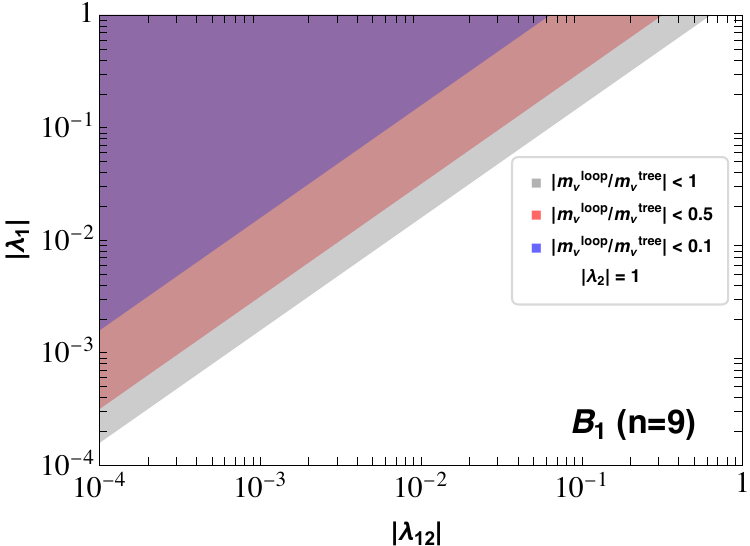}
    \includegraphics[width=0.495\linewidth]{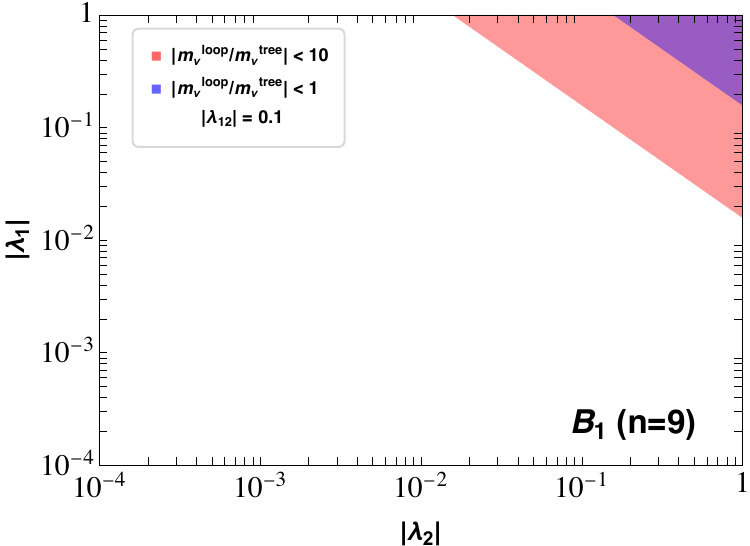}
    \caption{Same as Fig.~\ref{fig:A1} for Model ${\bf B_1}$, showing two different planes of the LNV quartic couplings.}
    \label{fig:B1}
\end{figure}
{In Figs.~\ref{fig:A1} and \ref{fig:B1} we show how the $\lambda$ couplings modify the $m_\nu^\text{loop}/m_\nu^{\text{tree}}$ ratio for Models ${\bf A_1}$ and ${\bf B_1}$, respectively.  We fix the new scalar mass to 400 GeV. The shaded region with oblique gray lines is  excluded due to the lower limits on pseudo-Nambu-Goldstone boson mass.
}
\section{Dimension-$6$ Effective Operators}
\label{sec:dim6}
At low energies, integrating-out the heavy fermion mediators, dimension-$6$ \emph{Derivative operators} are generated \cite{Herrero-Garcia:2016uab}. Unlike the dimension-$5$ Weinberg-like operators, these do not violate lepton number, but may violate lepton flavor. In particular, after SSB they modify couplings to SM gauge bosons, and yield a non-unitary PMNS matrix, $U \equiv (1-\kappa) U_L$, once canonically normalised neutrino kinetic terms are constructed \cite{deGouvea:2015euy,Blennow:2016jkn,Malinsky:2009df,Escrihuela:2016ube,Akhmedov:1995ip,Agarwalla:2021owd}. In the standard seesaw scenarios, such operators are discussed in Refs.~\cite{Abada:2007ux,Abada:2008ea}. For instance, for seesaw Type-I, the dimension-6 operator generated reads
\begin{equation} \label{eq:D6_SSI}
\mathcal{O}^{(0)}_6=\left(\overline{L}_\alpha \tilde{H}\right)i\slashed{\partial}\left( \tilde{H}^\dagger L_\beta\right)\,,
\end{equation}
with $C^{(0)}_6=\left( y_H M_N^{-2} y_H^\dagger\right)_{\alpha\beta}$. 

On general grounds, these dimension-$6$ operators are strongly suppressed because they depend on the same Yukawas as the dimension-$5$ ones responsible for neutrino masses and they are further suppressed by an additional power of the heavy mediator mass. Therefore, unless there exists a mechanism which decouples the dimension-$5$ (lepton-number-violating) and $6$ (lepton-number-conserving) WCs allowing a low-scale seesaw, as discussed in Section~\ref{sec:low}, the associated phenomenological signatures are suppressed. As shown in Section~\ref{sec:ISS_A1}, in the case Majorana mediators (Model $\bf A_1$), indeed an inverse seesaw scheme can be constructed. In this case, the larger Yukawas and smaller scales may yield significant contributions to dimension-$6$ operators. For the vector-like models (Models $\bf A_2$, $\bf B_i$), interestingly the product of Yukawas violates LN and enters in neutrino masses, as discussed in Section~\ref{sec:VL_low}. Therefore, the contributions to some dimension-$6$ operators, which only depend individually on either $y_1$ or $y_2$, can be significant.

In our \emph{genuine} models, dimension-$6$ operators involving the new scalar multiplets will be generated at low energies. Namely, for models that have $\Phi_1$ and a fermion $\mathcal{F}$, we have
\begin{equation} \label{eq:D6_phi1}
\mathcal{O}^{(1)}_6=\left(\overline{L}_\alpha {\Phi}_1\right)i\slashed{D}\left( {\Phi}_1^\dagger L_\beta\right)\,,
\end{equation}
with $C^{(1)}_6=\left( y_1 M_\mathcal{F}^{-2} y_1^\dagger\right)_{\alpha\beta}$. Here
\begin{equation}
    \slashed{D}=\slashed{\partial}-i\;\dfrac{g}{2}\Gamma^i \slashed{W}^i-i\;\dfrac{g^\prime}{2}Y\slashed{B}\
\end{equation}
is the covariant derivative, where $\Gamma^i$ are the $SU(2)$ generators of the representation of hypercharge $Y$ on which it acts, and the identity matrix is implicit in the hypercharge term. Similarly, for $\bf B_i$-models that also have a second multiplet, $\Phi_2$, identical operators $\mathcal{O}^{(2)}_6$ are generated, with the substitution $y_1\rightarrow y_2$ in $C^{(1)}_6\rightarrow C^{(2)}_6$. In principle, irrespective of neutrino masses and LNV, given the small VEVs, these will be further suppressed compared to $\mathcal{O}^{(0)}_6$. Therefore, the modifications to the gauge boson couplings are also expected to be generically smaller than in Type-I inverse seesaw. 

For the $\bf A_1$ model, $\mathcal{O}^{(1)}_6$ as in Eq.~\eqref{eq:D6_phi1} is generated with $C^{(1)}_6=( y_1 M_ \Sigma ^{-2}y_1^\dagger)_{\alpha\beta}$. For the $\bf A_{2}$ model, both $\mathcal{O}^{(0)}_6$ with just the SM Higgs can be built, as in Eq.~\eqref{eq:D6_SSI}, with coupling $C^{(0)}_6=( y_H M_\mathcal{F}^{-2}y_H^\dagger)_{\alpha\beta}$ \cite{Herrero-Garcia:2016uab}, and also $\mathcal{O}^{(1)}_6$ with the new multiplet $\Phi_1$ with $C^{(1)}_6=( y_1 M_\mathcal{F}^{-2}y_1^\dagger)_{\alpha\beta}$, as in Eq.~\eqref{eq:D6_phi1}. Unless there are large hierarchies in the Yukawas, the former one is expected to dominate. However, given that neutrino masses can be reproduced by taking a hierarchy of Yukawas and VEVs, for instance $y_1 v_1 \ll y_H v_H$,\footnote{The other hierarchy seems less natural, as it requires $y_H \ll 0.01 y_1$.} there may be significant effects from $C^{(0)}_6$, with $C^{(1)}_6\ll C^{(0)}_6$. As discussed in Section~\ref{sec:VL_low}, these hierarchical vector-like versions are the equivalent of inverse seesaw of hypercharge-less fermions. The key point is that the product of Yukawas and VEVs violates lepton number, while each of them individually does not and may contribute to $n=6$ operators.

Regarding the $\bf B_i$ models, where we add two different scalar multiplets with a vector-like mediator of mass $M_\mathcal{F}$, the integration of the heavy fields leads to both $\mathcal{O}^{(1)}_6$ and $\mathcal{O}^{(2)}_6$, with $C^{(1)}_6=\left( y_1 M_\mathcal{F}^{-2}y_1^\dagger\right)_{\alpha\beta}$ and $C^{(2)}_6=\left( y_2 M_\mathcal{F}^{-2} y_2^\dagger\right)_{\alpha\beta}$, respectively. Similarly to the case of the $\bf A_2$ model, neutrino masses can be reproduced by taking a hierarchy of Yukawas and VEVs, for instance $y_1 v_1\ll y_2 v_2$. Then, $C^{(2)}_6$ may be large, with $C^{(1)}_6\ll C^{(2)}_6$. Again, these are the equivalent of the inverse seesaw version of the Model $\bf A_1$. It is interesting to mention that dimension$-6$ operators may still be generated in $\mathbf{B_i}$ models if only one of the two new scalar multiplets take a VEV, differently from the neutrino masses case for which both non-zero VEVs are required.
\renewcommand{\arraystretch}{2}
\begin{table}[!htb]
\centering
{\small
\begin{tabular}{|c|c|c|c|c|c|c|}
\cline{2-7}
\multicolumn{1}{c|}{}&\multicolumn{3}{|c|}{$\mathcal{O}^{(0)}_6$}&\multicolumn{3}{|c|}{$\mathcal{O}^{(1)}_6$}\\
\cline{1-7}
\textbf{Model}&  $Z\;\nu_\alpha\nu_\beta$ & $Z\; e_\alpha e_\beta$&$W\;e\nu$&$Z\;\nu_\alpha\nu_\beta$ & $Z\; e_\alpha e_\beta$&$W\;e\nu$\\ \hline\hline
$\bf A_1$&\phantom{$-$}\ding{56}&\phantom{$-$}\ding{56}&\phantom{$-$}\ding{56}& $\dfrac{1}{2}$&\phantom{$-$}$-\dfrac{3}{4}$&\phantom{$-$}$\dfrac{17}{4}$\\ \hline
$\bf A_2$&\phantom{$-$}$1$&$-\dfrac{1}{2}$&\phantom{$-$}$\dfrac{1}{2}$&$-\dfrac{\sqrt{3}}{2}$&\phantom{$-$}$0$&$-\dfrac{\sqrt{3}}{6}$\\ \hline
\end{tabular}}
\caption{\label{tab:newGBcouplingsModels1H}Corrections to the $Z$- and $W$-boson couplings for the $\bf A_i$ models. For Model $\bf A_2$, the $\mathcal{O}^ {(1)}_6$ contribution is sub-leading with respect to the $\mathcal{O}^{(0)}_6$ one for $y_1 v_1 \ll y_H v$. The $Z$-boson couplings are in units of $\kappa_{\alpha\beta} e/(2c_W\;s_W)$, while $W$-boson couplings are in units of $\kappa_{\alpha\beta}  e/(2\sqrt{2}s_W)$. For instance, for Model $\bf A_2$, the value of $\kappa$ for $\mathcal{O}^{(0)}_6$ is $\kappa^{(0)}=\left(y_H M_\mathcal{F}^{-2} y_H^\dagger\right) v^2$ and for $\mathcal{O}^{(1)}_6$ is $\kappa^{(1)}=\left(y_1 M_\mathcal{F}^{-2} y_1^\dagger \right) v^2_1$.
Notice the different values to Ref.~\cite{Herrero-Garcia:2016uab} for $\bf A_2$, due to the different normalization that we used for $\braket{H}$.}
\end{table}
\renewcommand{\arraystretch}{1}

\renewcommand{\arraystretch}{2}
\begin{table}[!htb]
\centering
{\small
\begin{tabular}{|c|c|c|c|c|c|c|}
\cline{2-7}
\multicolumn{1}{c|}{}&\multicolumn{3}{|c|}{$\mathcal{O}^{(1)}_6$}&\multicolumn{3}{|c|}{$\mathcal{O}^{(2)}_6$}\\
\cline{1-7}
\textbf{Model}&  $Z\;\nu_\alpha\nu_\beta$ & $Z\; e_\alpha e_\beta$&$W\;e\nu$&$Z\;\nu_\alpha\nu_\beta$ & $Z\; e_\alpha e_\beta$&$W\;e\nu$\\ \hline\hline
$\bf B_1$&$-\dfrac{3}{4}$&\phantom{$-$}$\dfrac{1}{2}$&\phantom{$-$}$\dfrac{17}{4}$&\phantom{$-$}$\dfrac{3}{4}$& $-3$&$-\dfrac{3}{4}$\\ \hline
$\bf B_2$&\phantom{$-$}$0$&\phantom{$-$}$0$&\phantom{$-$}$4$&\phantom{$-$}$\dfrac{3}{4}$&$-\dfrac{1}{4}$&$-\dfrac{1}{4}$\\ \hline
$\bf B_3$&$-2$&\phantom{$-$}$0$&$-\dfrac{1}{2}$&$-1$&\phantom{$-$}$\dfrac{3}{4}$&$-\dfrac{5}{8}$\\ \hline
$\bf B_4$&\phantom{$-$}$\dfrac{3}{4}$&$-\dfrac{1}{4}$&$-\dfrac{1}{4}$ &\phantom{$-$}$0$&\phantom{$-$}$0$&\phantom{$-$}$\dfrac{1}{2}$\\ \hline
\end{tabular}}
\caption{Same as Table~\ref{tab:newGBcouplingsModels1H} for the $\bf B_i$ models. The $\mathcal{O}_6^{(2)}$ contribution will dominate w.r.t. that of $\mathcal{O}_6^{(1)}$ for $y_2 v_2 \gg y_1 v$}
\label{tab:newGBcouplingsModels2H}
\end{table}
\renewcommand{\arraystretch}{1}

When the scalars take VEVs, these operators modify the kinetic terms of the lepton doublets. Therefore, in order to obtain canonically normalized kinetic terms, we need to 
rescale the components of lepton doublet fields ($L\equiv (\nu, l)$): $\nu\rightarrow (1+\zeta_\nu\,\kappa)^{-1/2} \, \nu$ and $l \rightarrow (1+\zeta_l\,\kappa)^{-1/2}\, l$, where $\zeta_{\nu,l}$ are numerical factors (reported in Appendix~\ref{appb}) and $\kappa$ is a small parameter proportional to the dimension-$6$ couplings times the VEVs. For instance, for $\mathcal{O}^{(1)}_6$ with lepton flavors $\alpha \beta$, the $\kappa$ coupling is given by
\begin{equation}
\kappa^{(1)}_{\alpha \beta} = (C^{(1)}_6)_{\alpha\beta}\, v_1^2\,.
    \label{kappa}
\end{equation}
After an expansion in $\kappa$, we get modified couplings of the gauge bosons to the SM leptons. In Tables~\ref{tab:newGBcouplingsModels1H} and \ref{tab:newGBcouplingsModels2H} we present the  modifications to SM couplings for models $\bf A_i$ and $\bf B_i$, respectively. 
The corrections of $Z$-boson couplings to flavors $\alpha \beta$ are proportional to $\kappa_{\alpha \beta} e/(2s_Wc_W)$, while the $W$-boson couplings are proportional to $\kappa_{\alpha \beta} e/(2\sqrt{2}s_W)$.\footnote{Notice that the relation $g_W=g_{Z\nu}+g_{Ze}$ holds only for the Model $\bf A_2$ for the operator $\mathcal{O}^{(0)}_6$. This is because the SM Higgs doublet does not break the custodial symmetry. On the other hand, for the rest of the models, the relation does not hold since the new VEVs break custodial symmetry.} Note that $Z$-boson couplings are in all cases with left-handed lepton doublets.

\section{Phenomenology} \label{sec:pheno} 
In this section, we study the phenomenological signatures generated in the various models.

\subsection{Universality violation and non-unitarity of the leptonic mixing matrix}
\label{secdim6}
The dimension-$6$ operators modify the couplings of leptons to SM gauge bosons after SSB. In particular, they yield a non-unitary PMNS matrix, as well as $Z$-boson FCNC. For Model ${\bf A_2}$, also Higgs lepton flavor violating couplings are generated, see Ref.~\cite{Herrero-Garcia:2016uab} for details. The modified couplings due to $\mathcal{O}^{(0,1)}_6$ are given in Table \ref{tab:newGBcouplingsModels1H} for the $\bf A_i$ models, and those due to $\mathcal{O}^{(1,2)}_6$ are provided in Table \ref{tab:newGBcouplingsModels2H} for the $\bf B_i$ models. In the case of low-scale seesaws, these may give observable signals, as discussed in Section~\ref{sec:low}. 

The most constraining limits on the dimension-$6$ operators come from radiative LFV processes, $Z$-mediated tree-level LFV, non-universal decays, and from the non-unitarity of the PMNS mixing matrix. Although flavor-dependent, the typical constraints read
\begin{equation}
  \kappa^{(1)}_{\alpha\beta} \equiv c^{(1)}_{\alpha\beta} v^2_1=\left( y_1 M_F^{-2} y_1^\dagger\right)_{\alpha\beta} v_1^2 \lesssim \mathcal{O}(10^{-3})\,.
\end{equation}
In Table~\ref{tab:kll} we provide the limits on the diagonal (off-diagonal) entries of the $\kappa$ matrices of the different models for different flavors, from non-universal (LFV) $Z$-decays. The relevant expressions for the processes can be found in Section 3.2 of Ref.~\cite{Herrero-Garcia:2016uab}. As the matrix is Hermitian and positive-definite, the off-diagonal couplings are bounded by the diagonal ones, $|\kappa_{\ell \ell^\prime}|<\sqrt{\kappa_{\ell \ell} \kappa_{\ell^\prime \ell^\prime}}$. 

\begin{table}[!htb]
\centering
{\small
\begin{tabular}{|c|c|c|c|c|c|c|c|}
\cline{3-8}
\multicolumn{2}{c|}{}&\multicolumn{6}{|c|}{Upper limits}\\
\cline{1-8}
\hline
\textbf{Model}& $\mathcal{O}_6^{(i)}$ & $\kappa_{ee}$ &$\kappa_{\mu \mu}$ & $\kappa_{\tau \tau}$&$\kappa_{\tau \mu}$&$\kappa_{\tau e}$&$\kappa_{\mu e}$\\ 
\hline\hline
$\bf A_1$& $\mathcal{O}_6^{(1)}$ &$<0.0013$ & $<0.0028$ & $<0.0053$& $<0.0005$& $<0.0005$& $<1.3 \times 10^{-6}$\\ 
\hline
$\bf A_{2}$& $\mathcal{O}_6^{(0)}$ & $<0.0019$ & $<0.0042$ & $<0.0079$&$<0.0007$& $<0.0008$& $<2 \times 10^{-6}$\\ 
\hline
$\bf B_1$& $\mathcal{O}_6^{(1)}$ & $<0.0036$ & $<0.0042$ & $<0.0012$&$<0.0007$& $<0.0008$& $<2 \times 10^{-6}$\\ 
\hline
$\bf B_1$& $\mathcal{O}_6^{(2)}$ & $<0.0003$ & $<0.0007$ & $<0.0013$&$<0.0001$& $<0.0001$& $<3.3 \times 10^{-7}$\\ 
\hline
$\bf B_2$& $\mathcal{O}_6^{(2)}$ & $<0.0038$ & $<0.0084$ & $<0.0159$&$<0.0014$& $<0.0016$& $<4 \times 10^{-6}$\\ 
\hline
$\bf B_3$& $\mathcal{O}_6^{(2)}$ & $<0.0024$ & $<0.0028$ & $<0.0008$&$<0.0005$& $<0.0005$& $<1.3 \times 10^{-6}$\\ 
\hline
$\bf B_4$& $\mathcal{O}_6^{(1)}$ &$<0.0038$ & $<0.0084$ & $<0.0159$&$<0.0014$& $<0.0016$& $<4 \times 10^{-6}$\\  
\hline
\end{tabular}}
\caption{Limits on the dimension-6 operators from $Z$-boson mediated tree-level non-universal and LFV decays. For all models but ${\bf B_1}$, only one operator has a non-zero correction to $Z e_\alpha e_\beta$ and the corresponding limits are quoted. For ${\bf B_1}$, we derive the limits assuming only of them contributes while neglecting the other one.} \label{tab:kll} 
\end{table}

\subsection{Lepton Flavor Violation} \label{sec:LFV}

The new seesaw models generate LFV decays, which are absent in the SM. The experimental limits on these observables set bounds on the Yukawa couplings of the new scalars with charged leptons and the masses of the new particles. We focus on some of the most constraining processes, which are the radiative decays $l_\alpha \r l_\beta \gamma$. In our models, the analytical expression for ${\rm BR}(l_\alpha \r l_\beta \gamma)$ may be written as
\begin{align}\label{eq:lfv}
    {\rm BR}(l_\alpha \r l_\beta \gamma)=&\frac{3\alpha_{\rm em}}{64 \pi G_F^2}\,\bigg|\sum_{k=1}^{N_\Phi}\sum_{l=1}^{N_\Psi}y_{k}^{l{\beta^\ast}}y_{k}^{l\alpha}\sum_{j=1}^{n_{\Phi}}\sum_{i=1}^{n_\psi}\frac{c^2_{Q_{\Psi_i}Q_{\Phi_j}}}{M_{\Phi_j}^2}\left[Q_{\psi_i}f_\psi(r)+Q_{\Phi_j}f_{\Phi}(r)\right]\bigg|^2\nonumber\\
    &\times {\rm BR}(l_\alpha \r l_\beta \nu_\alpha \bar{\nu_b})\,,
\end{align}
where $r\equiv M_{\psi_i}^2/M_{\Phi_j}^2$, $\alpha_{\rm em}=e^2/(4\pi)$ is the fine structure constant and $G_F$ is Fermi's decay constant. Experimentally, the branching ratio ${\rm BR}(l_\alpha \r l_\beta \nu_\alpha \bar{\nu_b})$ is equal to $\{1,0.178,0.174\}$ for $\alpha\beta=\{\mu e, \tau e, \tau \mu\}$, respectively \cite{ParticleDataGroup:2022pth}. $N_{\Phi(\Psi)}$ denotes the number of new scalars (fermionic mediators, $\Psi=\Sigma,\mathcal{F}$) in the model, whereas $n_\Phi$ and $n_\psi$ denote the number of components in the scalar and fermion multiplets, respectively. $Q$ denotes the charges of the scalar and fermion components in the loop such that $Q_{l_\alpha}=Q_{\psi_i}-Q_{\Phi_j}$, and $c_{Q_{\psi_i}Q_{\Phi_j}}$ is the numerical coefficient of the term containing a scalar (fermion) component with charge $Q_{\Phi_j}$ ($Q_{\psi_i}$) in the expansion of the Yukawa Lagrangian in tensor notation. Finally, the loop functions are given by \cite{Lavoura:2003xp}
\begin{equation}
    \begin{aligned}
        f_\psi(x)&=\frac{2+3x-6x^2+x^3+6x\ln{x}}{6(1-x)^4}\approx \frac{1}{6x}({\rm for~}x \gg 1)\nonumber\,,\\
       f_\Phi(x)&=\frac{1-6x+3x^2+2x^3-6x^2 \ln{x}}{6(1-x)^4}\approx \frac{1}{3x}({\rm for~}x \gg 1)\,,
    \end{aligned}
\end{equation}
where in the last step we approximated them for large $x$.
We make the following assumptions for Eq.~\eqref{eq:lfv} to simplify the resulting expression:
\begin{itemize}
\item[\emph{i)}] The components within the multiplets are degenerate, i.e., $M_{\Phi_j}=M_\Phi\,,M_{\psi_i}=M_{\Psi}$, 
\item[\emph{ii)}] Whenever there are two or more new scalars or fermions in the model, we take their masses to be the same, i.e, $M_{\Phi_1}=M_{\Phi_2}=M_\Phi$, 
\item[\emph{iii)}] We ignore the mixing among the scalars, and among SM and heavy fermions, 
\item[\emph{iv)}] We consider only the contributions coming from BSM states in the loop. Further, in the limit $M_{\Sigma,\mathcal{F}} \gg M_\Phi$, the dependence on scalar masses is negligible, and the bounds can be expressed solely in terms of the fermion mediator mass only.
\end{itemize}

The most stringent bound comes from radiative muon decays, ${\rm BR}(\mu \r e \gamma)<4.2 \times 10^{-13}$ at 90\% CL \cite{MEG:2016leq}, which is much stronger than the radiative $\tau$-lepton decays, ${\rm BR}(\tau \r e \gamma)<3.3 \times 10^{-8}$ and ${\rm BR}(\tau \r \mu \gamma)<4.4 \times 10^{-8}$ at 90\% CL \cite{BaBar:2009hkt}. 

To illustrate the limits, let us consider the model $\bf A_1$ which contains one new scalar multiplet $\mathbf{4}_{-1/2}^S$ and one fermionic quintuplet $\mathbf{5}_0^F$. Given the charged lepton Yukawa Lagrangian expansion
\begin{equation}
   y_1\bar{L} \Phi \Sigma = y_1\bar{l}\left(\Sigma^{--}\Phi^+ -\frac{\sqrt{3}}{2}\Sigma^- \Phi^0+\frac{1}{\sqrt{2}}\Sigma^0 \Phi^- -\frac{1}{2}\Sigma^+\Phi^{--}\right)\,,
\end{equation}
we can write the branching ratio for $\mu \r e \gamma$ as
\begin{align}
    {\rm BR}(\mu \r e \gamma)=\frac{3\alpha_{\rm em}}{64 \pi G_F^2}\left|\frac{y_{1}^{{e^\ast}}y_{1}^{\mu}}{M_{\Phi}^2}\,  
    Z_{\Phi}^{\rm A1} \left(\frac{M_\Sigma^2}{M_\Phi^2}\right)
    \right|^2\,,
\end{align}
with
\begin{align}
    Z_{\Phi_1}^{\bf A_1} (x) &= \frac{1}{2}f_{\Phi}(x) -\frac{3}{4}f_\psi(x)-\left[2f_\psi(x) + f_\Phi(x)\right]+\frac{1}{4}\left[f_\psi(x)+2f_\Phi(x)\right]\,.
\end{align}
Imposing the current experimental upper bound on ${\rm BR}(\mu \r e \gamma)$, we get
\begin{equation}
\frac{|y_{1}^{{e^\ast}}y_{1}^{\mu}|}{({M_\Sigma}/{\rm TeV})^2}\lesssim 2 \times 10^{-4}\,.
\end{equation}
Similarly, bounds on various Yukawa combinations from other LFV decays can be calculated.
\begin{table}[!htb]
\begin{center}
{\small
\begin{tabular}{|c|c|c|c|c| }
\cline{3-5}
\multicolumn{1}{c}{}&\multicolumn{1}{c}{} &\multicolumn{3}{|c|}{Upper limits}\\
\cline{1-5}
\hline
 {\bf Model} & {\bf Yukawa combination} &\thead{$\alpha\beta=\mu e$}& \thead{$\alpha\beta=\tau e$}& \thead{$\alpha\beta=\tau\mu$}\\
\hline\hline
$\bf A_1$ & $|y_{1}^{{\beta^\ast}}y_{1}^{\alpha}|({\rm TeV}/{M_\Sigma})^2$&$<0.0002$ & $<0.13$ & $<0.16$\\
\hline
$\bf A_2$ & $|y_{1}^{{\beta^\ast}}y_{1}^{\alpha}|({\rm TeV}/{M_{\mathcal{F}}})^2$&$<0.0004$ & $<0.24$ & $<0.28$\\
\hline
$\bf B_1$ & $|y_{1}^{{\beta^\ast}}y_{1}^{\alpha}-0.5\,y_{2}^{{\beta^\ast}}y_{2}^{\alpha}|({\rm TeV}/{M_{\mathcal{F}}})^2$&$<0.0004$ & $<0.29$ & $<0.34$\\
\hline
$\bf B_2$ & $|y_{1}^{{\beta^\ast}}y_{1}^{\alpha}-50\,y_{2}^{{\beta^\ast}}y_{2}^{\alpha}|({\rm TeV}/{M_{\mathcal{F}}})^2$&$<0.0011$ & $<0.72$ & $<0.84$\\
\hline
$\bf B_3$ & $|y_{1}^{{\beta^\ast}}y_{1}^{\alpha}-2.12\,y_{2}^{{\beta^\ast}}y_{2}^{\alpha}|({\rm TeV}/{M_{\mathcal{F}}})^2$&$<0.0002$ & $<0.15$ & $<0.18$\\
\hline
$\bf B_4$ & $|y_{1}^{{\beta^\ast}}y_{1}^{\alpha}+6.6\,y_{2}^{{\beta^\ast}}y_{2}^{\alpha}|({\rm TeV}/{M_{\mathcal{F}}})^2$&$<0.0004$ & $<0.24$ & $<0.28$\\
\hline 
\end{tabular}}
\caption{$90\%$CL limits on combination of Yukawa couplings and fermion masses from lepton-flavor-violating radiative decays for the different models.}
\label{tab:lfvB}
\end{center}
\end{table}

In models of Type-${\bf B}$, which contain two new scalars, both Yukawas $y_{1}$ and $y_{2}$ in Eq.~\eqref{eq:lagrVL} enter the decay. Note that the Yukawa combination that is constrained from LFV decays is different from the one entering the expression of neutrino masses, where the product of both couplings enters. For example, consider the model $\bf B_1$ which contains the scalar multiplets $\mathbf{4}_{1/2}^S$, $\mathbf{4}_{-3/2}^S$ and a vector-like mediator $\mathbf{5}_0^F$. The relevant charged lepton Yukawa Lagrangian is
\begin{align}
    \bar{L} \Phi_1 \mathcal{F}_R &= \bar{l} \left(\mathcal{F}_R^{---}\Phi_1^{++}-\frac{\sqrt{3}}{2}\mathcal{F}_R^{--}\Phi_1^{+}-\frac{1}{\sqrt{2}}\mathcal{F}_R^{-}\Phi_1^0 -\frac{1}{2}\mathcal{F}_R^0 \Phi_1^{-}\right)\,,\nonumber\\
    \bar{L} \Phi_2 \mathcal{F}_L^c &=\bar{l} \left(-{\mathcal{F}_L^{+}}^c \Phi_2^0-\frac{\sqrt{3}}{2}{\mathcal{F}_L^{0}}^c \Phi_2^{-}-\frac{1}{\sqrt{2}}{\mathcal{F}_L^{-}}^c \Phi_2^{--} -\frac{1}{2}{\mathcal{F}_L^{--}}^c \Phi_2^{---}\right)\,,
\end{align}
which gives
\begin{align}
    &{\rm BR}(l_\alpha \r l_\beta \gamma)=\frac{3\alpha_{\rm em}}{64 \pi G_F^2}\,\bigg|\frac{y_{1}^{\beta\ast}y_{1}^{\alpha}}{M_{\Phi_1}^2}Z_{\Phi_1}^{\rm B1} \left(\frac{M_\mathcal{F}^2}{M_{\Phi_1}^2}\right)+\frac{y_{2}^{\beta\ast}y_{2}^{\alpha}}{M_{\Phi_2}^2}Z_{\Phi_2}^{\rm B1}\left(\frac{M_\mathcal{F}^2}{M_{\Phi_2}^2}\right)\bigg|^2\times {\rm BR}(l_\alpha \r l_\beta \nu_\alpha \bar{\nu_b})\,,
\end{align}
with
\begin{align}
    Z_{\Phi_1}^{\bf B_1} (x) &= \left\{-[3f_\psi(x)+2f_\Phi (x)]-\frac{3}{4}[2f_\psi(x)+f_\Phi(x)]-\frac{1}{2}[f_\psi(x)]+\frac{1}{4}[f_\Phi(x)]\right\}\,,\nonumber\\
    Z_{\Phi_2}^{\bf B_1} (x)&= \left\{-[f_{\psi}(x)]+\frac{3}{4}[f_\Phi(x)]+\frac{1}{2}[f_\psi(x)+2f_\Phi(x)]+\frac{1}{4}[2f_\psi(x)+3f_\Phi(x)]\right\}\,.
\end{align}
Taking $x\gg1$, we can simplify the above expressions as $Z_{\Phi_1}^{\rm B1}(x) \approx -5/(6x)$ and $Z_{\Phi_2}^{\rm B1}(x) \approx 5/(12x)$. Then, the bound from BR$(\mu \r e \gamma)$ reads
\begin{equation}
\frac{|y_{1}^{e\ast}y_{1}^\mu-0.5\,y_{2}^{e\ast}y_{2}^\mu|}{(M_\mathcal{F}/{\rm TeV})^2} < 4.3 \times 10^{-4}\,.
\end{equation}
The bounds on the Yukawa combinations from LFV decays for all models are summarised in Table~\ref{tab:lfvB}. Let us also mention that the Yukawa structure can be determined in terms of neutrino parameters using the Casas-Ibarra parameterization for Majorana-like mediators \cite{Casas:2001sr} (Model ${\bf A_1}$), whereas for vector-like mediators in ${\bf B}$-type models, the approach used in the Generalized Scotogenic Model can be followed \cite{Hagedorn:2018spx}. This would be useful to do a parameter scan of the models, which is beyond the scope of this work.

Moreover, one can also study other LFV processes such as $l_\alpha \r 3l_\beta$ and $\mu-e$ conversion in nuclei in our models \cite{Ren:2011mh,Toma:2013zsa,Felkl:2021qdn}. The experimental sensitivities for these processes are expected to improve by four orders of magnitude in the coming years \cite{Perrevoort:2023jry} and it would be worth studying them in detail for a specific model. The former process can receive contributions from penguins and box-type diagrams, however, for small values of Yukawas (as seen in the analysis above), the box-type contribution will be subdominant, and using \textit{dipole dominance approximation}, we can write \cite{Hagedorn:2018spx}
\begin{align}
    {\rm BR}(\mu \r 3e) &\approx \frac{\alpha_{\rm em}}{8\pi}\left(\frac{16}{3}\ln\frac{m_\mu}{m_e}-\frac{22}{3}\right) \times {\rm BR}(\mu \r e \gamma)\nonumber\\
    &\approx 0.006 \times {\rm BR}(\mu \r e \gamma)\,.
\end{align}
The above approximation can be used to further constrain the Yukawas, when new data becomes available in the future. 

Finally, there also exist implications for charged lepton electric dipole moments (EDM) in our models, which are generated at the two-loop level. The electric dipole moments for the minimal scotogenic model \cite{Ma:2006km} were calculated in Ref.~\cite{Abada:2018zra}. The sensitivity to the muon EDM is expected to improve by more than three orders of magnitude  at the muEDM experiment based in PSI \cite{Schmidt-Wellenburg:2023aga}.  It is possible to construct scotogenic-like models from our models as well, as we discuss below. 

\section{(Generalised) Scotogenic-like models} \label{sec:scot}

Models where neutrino masses arise purely at the loop level are interesting because a viable dark matter (DM) candidate may be easily embedded, see Ref.~\cite{Arbelaez:2022ejo,Restrepo:2013aga,Cai:2017jrq}. Indeed, if the new scalar and Majorana (vector-like) fermion multiplets transform under a $Z_2$ ($U(1)$) dark symmetry, the lightest of them, if neutral, may be a viable DM candidate. If the particle has $Y=0$, it may be allowed by direct detection (DD) constraints if the Higgs portal couplings in the potential are small enough. For the $Y\neq0$ scenarios, DD excludes them, unless the DM candidate is a CP-even (odd) scalar with a sufficiently-large mass splitting (i.e., $\Delta m_\Phi \gtrsim \mathcal{O}\,(\text{MeV})$) with its CP-odd (even) counterpart \cite{Cirelli:2005uq,Hambye:2009pw}. Another way to have a viable scalar DM candidate from a multiplet with $Y\neq0$ is to mix the neutral component with a real scalar singlet, as in the ScotoSinglet Model~\cite{Beniwal:2020hjc}. These (generalised) Scotogenic-like models are:
\begin{itemize}
\item Scotogenic-like Model $\bf A^\prime_1$: with a $Z_2$ symmetry, such that $\Phi\rightarrow -\Phi$ and $\Sigma \rightarrow - \Sigma$. Imposing the $Z_2$ symmetry kills the terms involving only a single scalar multiplet in the scalar potential.
\item Generalised Scotogenic-like Model $\bf B^\prime_i$: with a new global $U(1)$ symmetry, such that only $\Phi_1$, $\Phi_2$ and $\mathcal{F}$ transform non-trivially, for example: $q_{\Phi_1}=1, q_{\Phi_2}=-1$ and $q_\mathcal{F}=-1$, the generalised scotogenic equivalent term will be $HH \Phi_1 \Phi_2$. Note that the same term can lead to the needed mass splitting among the real and imaginary parts of the scalar multiplets with $Y\neq 0$. It can be seen that this new global symmetry can be identified as the $U(1)_X$ accidental symmetry of the scalar potential involving two new scalar multiplets, see Table~3 of Ref.~\cite{Giarnetti:2023dcr}.
\end{itemize}
Note that as the Model $\bf A_2$ is the only model that includes the SM Higgs to generate neutrino masses radiatively via the term $HHH\Phi^\ast$, neither $Z_2$ nor $U(1)$ symmetry can be imposed to accommodate a DM candidate. On the other hand, given the large representations of the new particles in models ${\bf B_3^\prime}$ and ${\bf B_4^\prime}$, the global U(1) symmetry of the latter scenarios is an accidental symmetry at the renormalisable level, see also Ref.~\cite{Arbelaez:2022ejo}.
\begin{table}[!htb]
\centering
{\small
\begin{tabular}{|c|c|c|c|c|c|}
\hline
\textbf{Model}& \textbf{New fields} &\textbf{Sym.} &\textbf{DM candidates} & \textbf{DM Mass (TeV)} \\ 
\hline\hline
$\bf A^\prime_1$&$\Phi_1=4_{-1/2}^S\,,\Sigma=5_0^F$ &$Z_2$&$4_{-1/2}^S$, $5_0^F$& $M_{\Phi_1}\approx 3.2,M_{\Sigma}\approx 10$\\ 
\hline
$\bf A^\prime_{2}$&$\Phi_1=4_{-3/2}^S\,,\mathcal{F}=3_{-1}^F$&$-$&$-$&$-$  \\ 
\hline
$\bf B^\prime_1$&$\Phi_1=4_{1/2}^S\,,\Phi_2=4_{-3/2}^S\,,\mathcal{F}=5_{-1}^F$&$U(1)$&$4_{1/2}^S\,,4_{-3/2}^S$&$M_{\Phi_1}\approx 3.2,M_{\Phi_2}\approx 3.5$\\ 
\hline
$\bf B^\prime_2$&$\Phi_1=3_{0}^S\,,\Phi_2=5_{-1}^S\,,\mathcal{F}=4_{-1/2}^F$&$U(1)$&$3_0^S\,,5_{-1}^S$&$M_{\Phi_1}\approx 2.5,M_{\Phi_2}\approx 3.4$\\ 
\hline
$\bf B^\prime_3$&$\Phi_1=5_{-2}^S\,,\Phi_2=5_{1}^S\,,\mathcal{F}=4_{3/2}^F$&$U(1)$&$5_{-2}^S\,,5_{1}^S$&$M_{\Phi_1}\approx 3.9,M_{\Phi_2}\approx 3.4$\\
\hline
$\bf B^\prime_4$&$\Phi_1=5_{-1}^S\,,\Phi_2=5_{0}^S\,,\mathcal{F}=4_{1/2}^F$&$U(1)$&$5_{-1}^S\,,5_{0}^S$&$M_{\Phi_1}\approx 3.4,M_{\Phi_2}\approx 9.4$\\ 
\hline
\end{tabular}}
\caption{List of \emph{(Generalised) Scotogenic}-like models which generate neutrino masses at one loop. We give the stabilising symmetry in the third column and the possible DM candidates in the fourth column. The mass for the DM candidate that reproduces the observed relic abundance is listed in the last column, including non-perturbative effects for the $Y=0$ candidates \cite{Cirelli:2007xd}.} \label{tab:listScot} 
\end{table}

In Table~\ref{tab:listScot} we summarise the \emph{Scotogenic}-like models, labelled by $\bf A_i ^\prime$ and $\bf B_i ^\prime$ outlining their dark symmetry and their potentially-viable DM candidates. We also indicate the corresponding DM mass required to reproduce the DM relic abundance, $\Omega_{\rm DM}h^2 = 0.12 \pm 0.0012$ where we assume that the neutral component of only one of the candidates makes up the total relic abundance. We have computed the masses following the approach of Minimal Dark Matter, see Refs.~\cite{Cirelli:2005uq,Cirelli:2007xd}, where all co-annihilations (both $s$-wave and $p$-wave) of the DM component along with the relevant RGE corrections have been taken into account in the perturbative approximation. Furthermore, for candidates with $Y=0$, the non-perturbative Sommerfeld correction becomes quite important, especially for multiplets belong to higher $SU(2)$ representation, which enhances the annihilation cross-section compared to the perturbative approach, thus leading to larger DM masses in order to obtain the correct relic abundance. Therefore, the masses indicated for $3_0^S, 5_0^S$ and $5_0^F$ are obtained after including these non-perturbative effects \cite{Cirelli:2007xd}.\footnote{These computations were done at LO, the NLO electroweak potential computations can be found in Ref.~\cite{Urban:2021cdu}.} The precise computation of the DM masses corresponding to each multiplet after taking all these effects into account and a detailed analysis of direct and indirect detection limits would be interesting, but it is quite involved and beyond the scope of the present work.

\section{Conclusions} \label{sec:conc} 

The three standard seesaw mechanisms explaining small neutrino masses involve the SM Higgs doublet. In the presence of new scalar and fermion multiplets, several other options emerge. One of the main advantages of these higher-representation seesaw models is that the vacuum expectation values (VEVs) of the new scalars, and consequently neutrino masses, are suppressed for reasons unrelated to the violation of lepton number: namely, the constraints stemming from the $\rho$ parameter due to the violation of custodial symmetry ~\cite{Giarnetti:2023dcr}.

In this work, we have studied possible seesaw scenarios that do not involve the states present in the standard seesaws, neither the same fermions nor contributions that solely depend on the SM Higgs doublet VEV. These requirements narrow down the options to $2,(4)$ models featuring $1, (2)$ new scalar multiplet(s) up to quintuplet $SU(2)$ representations. Only one involves a Majorana fermion mediator, while the rest involve vector-like mediators. These models are interesting due to their richer phenomenology compared to the standard Type-I seesaw scenario. In Ref.~\cite{Giarnetti:2023dcr} we have studied the scalar phenomenology, while here we focus on the fermion phenomenology.

Furthermore, low-energy versions can be constructed in all cases, with some small parameter controlling the violation of lepton number, such as the Majorana mass $\mu_L$ in the inverse seesaw variant of Model ${\bf A_1}$ (see Eq.~\eqref{eq:ISS_Maj}), or a Yukawa coupling for the other models featuring vector-like fermions, for instance, $y_1$ in Eq.~\eqref{eq:low_vec}. In these low-energy versions, the effects of dimension-$6$ operators may be significant. In fact, the latter may generate substantial contributions to lepton flavor-violating observables, as well as to the non-unitarity of the PMNS mixing matrix and non-universal decays of the $Z$-boson.

In the considered seesaw models, one-loop corrections to neutrino masses are also present. Specifically, in models ${\bf B_{3,4}}$, when TeV-scale scalars and order one couplings are involved, these loop contributions dominate. Additionally, it is interesting to note that, in certain instances, \emph{(Generalized) Scotogenic-like} models capable of generating neutrino masses solely at one loop can be formulated. For scenarios where the fermion is Majorana (vector-like), a new $Z_2$ (global U(1)) symmetry must be imposed. In some of the resulting scenarios, a potentially viable DM candidate naturally emerges.

Let us briefly consider the possibility of leptogenesis in these scenarios. As the scalar multiplets acquire smaller VEVs, one might expect the lower bound on the mediator mass to roughly scale by the ratio of $v_{i}/v$ compared to the lower bound of $\mathcal{O}(10^9)$ GeV for fermion singlets \cite{Davidson:2002qv}. However, these new multiplets possess gauge interactions that deplete their population in the early universe, consequently suppressing the efficiency of leptogenesis. For instance, in the case of fermion triplets, the required mass of the lightest fermion triplet is estimated to be around $\mathcal{O}(10^{10})$ GeV \cite{Hambye:2012fh}, assuming a hierarchical spectrum of the added fermion mediators to the model. In a related study \cite{Vatsyayan:2022rth}, it was demonstrated that even if the fermion triplet couples to a new scalar doublet obtaining a VEV significantly lower than the SM Higgs, the scale can only be lowered to $10^6$ GeV (including flavor effects). Consequently, we anticipate that for multiplets belonging to higher $SU(2)$ representations, the constraints become more stringent due to their multiply-charged states and stronger gauge interactions. This suggests a necessity for heavy fermion masses of approximately $\mathcal{O}(10^{10})$ GeV or higher for successful leptogenesis in such scenarios. Therefore, accommodating low-scale leptogenesis within these models appears to be challenging.

In conclusion, while somewhat less minimal compared to standard seesaws, the models analysed in this study are intriguing and offer a rich phenomenology, particularly in their low-energy versions. Should large $SU(2)$ representations be discovered, this research could offer valuable insights into their impact on the generation of light neutrino masses.

\vspace{0.3cm}

\acknowledgments

DV would like to thank Srubabati Goswami for useful discussions.
All Feynman diagrams were generated using the Ti\textit{k}Z-Feynman package for \LaTeX~\cite{Ellis:2016jkw}. JHG and DV are partially supported by the ``Generalitat Valenciana'' through the GenT Excellence Program (CIDEGENT\slash 2020\slash 020) and by the Spanish ``Agencia Estatal de Investigación'', MICINN\slash AEI (10.13039\slash 501100011033) grants PID2020-113334GB-I00 and PID2020-113644GB-I00. JHG is also supported by the ``Consolidación Investigadora'' Grant CNS2022-135592 funded by the Spanish ``Agencia Estatal de Investigación'', MICINN\slash AEI (10.13039\slash 501100011033) and by ``European Union NextGenerationEU/PRTR''.

\appendix

\section{Naturalness considerations} \label{sec:natHiggs} 

Since the new scalars take VEVs, their masses are bounded from above, i.e., they should be below the  TeV scale. On the other hand, being scalars, there is the well-known hierarchy problem due to the presence of heavy scales, in this case, the heavy fermion masses $m_\Psi$. In seesaw Type-I, the one-loop correction to the Higgs boson self-energy from light and heavy sterile neutrinos ($N_R$) in the loop reads
\begin{equation}
    \delta_{\rm \overline{MS}}\, m^2_h = \frac{m_\nu m_{N_R}^3}{4 \pi v^2} \left[ \log\left(\frac{m^2_{N_R}}{\mu^2}-1\right)\right]\,,
\end{equation}
in the $\overline{\rm MS}$ scheme \cite{Vissani:1997ys, Farina:2013mla, Casas:2004gh}. It implies that the masses of heavy sterile neutrinos should be below $m_{N_R}\lesssim 10^7$ GeV if there is no fine-tuning with the bare Higgs mass, i.e., $\delta_{\rm \overline{MS}}\, m^2_h/m_h^2<1$.

In our scenarios, both the neutral and the singly-charged fermions run in the one-loop self-energy. Taking the multiplets $S$ and $\Psi$ to be degenerate, the contribution to the new scalar masses  at one loop is roughly
\begin{equation}
    \delta_{\rm \overline{MS}}\, m^2_S \simeq \mathcal{O}(1)\,\frac{m_\nu m_\Psi^3}{4 \pi v_i^2} \left[ \log\left(\frac{m^2_\Psi}{\mu^2}-1\right)\right]\,.
\end{equation}
For TeV mass scalars, taking the new VEVs $v_i\lesssim \text{GeV}$ ($\lesssim\,\text{MeV}$), we obtain $m_\Psi \lesssim 10^6\,(10^4)$ GeV. However, in the case of non-singlet $SU(2)$ heavy fermions, two-loop contributions to the self-energy dominate, due to the fact that the new multiplets have gauge interactions~\cite{Farina:2013mla}, see also Ref.~\cite{Herrero-Garcia:2019czj}. The upper limit on the mass is much stronger, of order a few TeVs, similar to those of Type-II and Type-III seesaws \cite{Farina:2013mla}.

\renewcommand{\arraystretch}{2}
\begin{table}[!htb]
\centering
{\small
\begin{tabular}{|c|c|c|c|c|c|c|}
\cline{2-7}
\multicolumn{1}{c|}{}&\multicolumn{2}{|c|}{$\mathcal{O}^{(0)}_6$}&\multicolumn{2}{|c|}{$\mathcal{O}^{(1)}_6$}&\multicolumn{2}{|c|}{$\mathcal{O}^{(2)}_6$}\\
\cline{1-7}
\textbf{Model}&  $\zeta_\nu$ & $\zeta_l$&$\zeta_\nu$ & $\zeta_l$&$\zeta_\nu$ & $\zeta_l$\\ \hline\hline
$\bf A_1$&\phantom{$-$}\ding{56}&\phantom{$-$}\ding{56}&$\dfrac{\sqrt{2}}{2}$& $\dfrac{3\sqrt{2}}{4}$&\phantom{$-$}\ding{56}&\phantom{$-$}\ding{56}\\ \hline
$\bf A_2$&$-\sqrt{2}$&$-\dfrac{1}{\sqrt{2}}$&$-\dfrac{1}{\sqrt{6}}$&$0$&\phantom{$-$}\ding{56}&\phantom{$-$}\ding{56}\\ \hline
$\bf B_1$&\phantom{$-$}\ding{56}&\phantom{$-$}\ding{56}&\phantom{$-$}$\dfrac{3\sqrt{2}}{4}$&\phantom{$-$} $\dfrac{1}{\sqrt{2}}$&\phantom{$-$}$\dfrac{\sqrt{2}}{4}$&\phantom{$-$}$\dfrac{4}{\sqrt{2}}$\\ \hline
$\bf B_2$&\phantom{$-$}\ding{56}&\phantom{$-$}\ding{56}&$-\dfrac{4}{3\sqrt{2}}$& $-\dfrac{4}{3\sqrt{2}}$&\phantom{$-$}$\dfrac{3}{4\sqrt{2}}$&\phantom{$-$}$\dfrac{1}{4\sqrt{2}}$\\ \hline
$\bf B_3$&\phantom{$-$}\ding{56}&\phantom{$-$}\ding{56}&$-\dfrac{1}{\sqrt{2}}$&\phantom{$-$}0&\phantom{$-$}$\dfrac{1}{\sqrt{2}}$&\phantom{$-$}$\dfrac{3}{4\sqrt{2}}$\\ \hline
$\bf B_4$&\phantom{$-$}\ding{56}&\phantom{$-$}\ding{56}&\phantom{$-$}$\dfrac{3}{4\sqrt{2}}$&\phantom{$-$}$\dfrac{1}{4\sqrt{2}}$&$-\dfrac{1}{\sqrt{2}}$&$-\dfrac{1}{\sqrt{2}}$\\ \hline
\end{tabular}}
\caption{\label{tab:zetas}Numerical factors $\zeta$ corresponding to $\mathcal{O}^{(i)}_6$ used for rescaling the lepton doublet fields.}
\end{table}
\renewcommand{\arraystretch}{1}

\section{Canonical normalisation of the lepton kinetic terms}\label{appb}

As discussed in Section~\ref{sec:dim6}, after EW SSB, the dimension$-6$ operators modify the kinetic term of the lepton doublets as follows
\begin{equation}
    \mathcal{L}^{\rm kin}_{\rm SM}+\sum_i \frac{\kappa^{(i)}}{v_i^2}\mathcal{O}^{(i)}_6\,\quad \text{with } i=0,1,2\,,
\end{equation}
where $\kappa^{(i)}=C^{(i)}_6\,v_i^2$ is defined in Eq.~\eqref{kappa}. 

In order to obtain a canonically normalized kinetic term, we need to re-scale the left-handed lepton doublet fields: $\nu\rightarrow (1+\zeta_\nu\,\kappa)^{-1/2} \, \nu$ and $l \rightarrow (1+\zeta_l\,\kappa)^{-1/2}\, l$, where $\zeta_{\nu,l}$. We report for each operator $\mathcal{O}^{(i)}_6$ ($i=1,2,3$), we report in Table~\ref{tab:zetas} the numerical factors $\zeta_{\nu,l}$.

\bibliographystyle{JHEP}
\bibliography{Weinbergs}
\end{document}